\PassOptionsToPackage{unicode}{hyperref}
\PassOptionsToPackage{hyphens}{url}
\PassOptionsToPackage{dvipsnames,svgnames,x11names}{xcolor}

\documentclass[
  12pt]{article}

\usepackage{amsmath,amssymb}
\usepackage{iftex}
\ifPDFTeX
  \usepackage[T1]{fontenc}
  \usepackage[utf8]{inputenc}
  \usepackage{textcomp} 
\else 
  \usepackage{unicode-math}
  \defaultfontfeatures{Scale=MatchLowercase}
  \defaultfontfeatures[\rmfamily]{Ligatures=TeX,Scale=1}
\fi
\usepackage{lmodern}

\ifPDFTeX\else  
    
\fi

\IfFileExists{upquote.sty}{\usepackage{upquote}}{}
\IfFileExists{microtype.sty}{
  \usepackage[]{microtype}
  \UseMicrotypeSet[protrusion]{basicmath} 
}{}
\makeatletter
\@ifundefined{KOMAClassName}{
  \IfFileExists{parskip.sty}{
    \usepackage{parskip}
  }{
    \setlength{\parindent}{0pt}
    \setlength{\parskip}{6pt plus 2pt minus 1pt}}
}{
  \KOMAoptions{parskip=half}}
\makeatother
\usepackage{xcolor}
\makeatletter
\ifx\paragraph\undefined\else
  \let\oldparagraph\paragraph
  \renewcommand{\paragraph}{
    \@ifstar
      \xxxParagraphStar
      \xxxParagraphNoStar
  }
  \newcommand{\xxxParagraphStar}[1]{\oldparagraph*{#1}\mbox{}}
  \newcommand{\xxxParagraphNoStar}[1]{\oldparagraph{#1}\mbox{}}
\fi
\ifx\subparagraph\undefined\else
  \let\oldsubparagraph\subparagraph
  \renewcommand{\subparagraph}{
    \@ifstar
      \xxxSubParagraphStar
      \xxxSubParagraphNoStar
  }
  \newcommand{\xxxSubParagraphStar}[1]{\oldsubparagraph*{#1}\mbox{}}
  \newcommand{\xxxSubParagraphNoStar}[1]{\oldsubparagraph{#1}\mbox{}}
\fi
\makeatother

\usepackage{longtable,booktabs,array}
\usepackage{calc} 
\usepackage{etoolbox}
\usepackage{multirow}
\makeatletter
\patchcmd\longtable{\par}{\if@noskipsec\mbox{}\fi\par}{}{}
\makeatother

\IfFileExists{footnotehyper.sty}{\usepackage{footnotehyper}}{\usepackage{footnote}}
\makesavenoteenv{longtable}
\usepackage{graphicx}
\makeatletter
\def\maxwidth{\ifdim\Gin@nat@width>\linewidth\linewidth\else\Gin@nat@width\fi}
\def\maxheight{\ifdim\Gin@nat@height>\textheight\textheight\else\Gin@nat@height\fi}
\makeatother

\setkeys{Gin}{width=\maxwidth,height=\maxheight,keepaspectratio}

\makeatletter
\def\fps@figure{htbp}
\makeatother

\makeatletter
\@ifpackageloaded{caption}{}{\usepackage{caption}}
\AtBeginDocument{
\ifdefined\contentsname
  \renewcommand*\contentsname{Table of contents}
\else
  \newcommand\contentsname{Table of contents}
\fi
\ifdefined\listfigurename
  \renewcommand*\listfigurename{List of Figures}
\else
  \newcommand\listfigurename{List of Figures}
\fi
\ifdefined\listtablename
  \renewcommand*\listtablename{List of Tables}
\else
  \newcommand\listtablename{List of Tables}
\fi
\ifdefined\figurename
  \renewcommand*\figurename{Figure}
\else
  \newcommand\figurename{Figure}
\fi
\ifdefined\tablename
  \renewcommand*\tablename{Table}
\else
  \newcommand\tablename{Table}
\fi
}
\@ifpackageloaded{float}{}{\usepackage{float}}
\floatstyle{ruled}
\@ifundefined{c@chapter}{\newfloat{codelisting}{h}{lop}}{\newfloat{codelisting}{h}{lop}[chapter]}
\floatname{codelisting}{Listing}

\makeatother
\makeatletter
\@ifpackageloaded{caption}{}{\usepackage{caption}}
\@ifpackageloaded{subcaption}{}{\usepackage{subcaption}}
\makeatother

\ifLuaTeX
  \usepackage{selnolig}  
\fi
\usepackage[]{natbib}
\usepackage{bookmark}

\IfFileExists{xurl.sty}{\usepackage{xurl}}{} 
\hypersetup{
  pdftitle={Title},
  pdfauthor={Author 1; Author 2},
  pdfkeywords={3 to 6 keywords, that do not appear in the title},
  colorlinks=true,
  linkcolor={blue},
  filecolor={Maroon},
  citecolor={Blue},
  urlcolor={Blue},
  pdfcreator={LaTeX via pandoc}}

\usepackage{bm}

\newcommand{\bs}{\textbf{s}}

\DeclareMathOperator{\dist}{dist}
\DeclareMathOperator{\elev}{elev}

\newcommand{\anon}{1}

\begin{document}

\def\spacingset#1{\renewcommand{\baselinestretch}
{#1}\small\normalsize} \spacingset{1}

\if1\anon
{
  \title{\bf Spatially varying quantile downscaling of summer temperature from atmospheric circulation and long-term trends over peninsular Spain}
  
  \author{Juan Marcen-Gutierrez\textsuperscript{1,}\thanks{Corresponding author: \texttt{jmarcen@unizar.es}}, and
          Jorge Castillo-Mateo\textsuperscript{1}  \\
          \small\textsuperscript{1}Department of Statistical Methods and IUMA, University of Zaragoza }
  \date{}
  \maketitle
} \fi

\if0\anon
{
  \bigskip
  \bigskip
  \bigskip
  \begin{center}
    {\LARGE\bf Title}
\end{center}
  \medskip
} \fi

\bigskip
\begin{abstract}
Over recent decades, high temperatures have intensified worldwide, with impacts on human health, ecosystems, and the economy. Understanding their spatial evolution requires accounting for long-term temperature changes and the role of atmospheric circulation. Quantile regression provides a flexible framework for characterizing changes across the temperature distribution, with relevance to extremes. We analyze the spatio-temporal evolution of the 0.50 and 0.95 quantiles of daily maximum temperature during summer (June--August) over peninsular Spain from 1960 to 2023. We develop a Bayesian spatially varying quantile regression model for observations from 40 meteorological stations, using ERA5 geopotential fields at tropospheric levels to characterize atmospheric circulation, together with long-term temporal trends capturing the background climate change signal. The spatially varying coefficients allow the sensitivity of temperature to atmospheric circulation to differ across the region, linking synoptic dynamics with local temperature variability. The model separates changes associated with the dynamical atmospheric state from the long-term thermodynamic warming signal, while allowing their effects to vary across the temperature distribution. Results show that atmospheric circulation and long-term trends provide comparable contributions to the observed warming. Significant increases in median and upper-quantile temperatures are found across much of peninsular Spain, with pronounced heterogeneity across months and space, followed by quantiles.
\end{abstract}

\noindent
{\it Keywords:} Bayesian hierarchical modeling, Gaussian process, Markov chain Monte Carlo, quantile regression, spatially varying coefficient, statistical downscaling
\vfill

\newpage
\spacingset{1.4} 

\section{Introduction}

The Mediterranean region, and in particular peninsular Spain, is undergoing an increase in temperatures that outpaces the global average \citep{lionello2018, urdiales2023}, differing in space and between seasons \citep{sandonis2021}. The increase in temperatures affects the frequency and intensity of extreme weather events, such as heatwaves \citep{lorenzo2021}, which cause great losses in human lives, mainly due to cardiorespiratory diseases \citep{cheng2019}; environmental damage; and economic losses \citep{cardil2015}.
In particular, the heatwaves suffered in Spain between June and August in $2003$, $2022$, and $2023$ claimed thousands of lives \citep{gil2024, barba2026}. In response, in 2004, Spanish health authorities established heatwave thresholds based on the $0.95$ quantile of maximum summer temperatures over $1980$--$2003$ \citep{gil2024}. 

In this context of concern about extreme heat, an effort in order to study the evolution of temperatures, particularly the evolution of extremes, has been made in Spain \citep{clivar2025}. An important part of the analysis of extremes is based on the definition of high quantiles, either because of their intrinsic interest  \citep{castillo2023}, or because they allow for the definition of exceedance over thresholds \citep{cebrian2022}. In the case of climatic variables, such as daily maximum temperatures, quantiles provide a complete view of the thermal behavior of a region. Consequently, modeling of upper-tail temperature quantiles has gained growing interest for evaluating the spatio-temporal evolution of extreme heat events. 

The analysis of a distribution given by observed data through quantile regression (QR) supposes a better capture of heterogeneous changes through quantiles with respect to the traditional statistical methods, which usually focus on the study of the mean \citep{koenker2005}. Additionally, these models present robust properties through the use of asymmetric loss functions \citep{huber2011}. QR may be addressed in two different ways. The first one follows the original idea of \cite{koenker1978}, in which each quantile is modeled separately. In a Bayesian framework, the use of asymmetric Laplace (AL) errors \citep{yu2001} enables the use of Markov chain Monte Carlo \citep[MCMC;][]{kozumi2011} for estimation purposes. Secondly, quantiles may be modeled jointly \citep{yang2017, chen2021, castillo2024}. This methodology avoids quantile crossing; however, its high complexity of implementation and restrictions in covariates and spatial effects found in the literature make it unfeasible when working with large datasets. 

In the climate field, QR has gained recent popularity due to its capability of identifying how different parts of the distribution respond to different geographical factors. 
\cite{gao2017} study temporal trends and spatial patterns of changes in extreme temperatures through QR models and spatial clustering techniques.
\cite{kalisa2021} employ local QR models to investigate the spatio-temporal variability of wet and dry conditions over East Africa.
\cite{solaimani2024} compare frequentist and Bayesian QR models when studying the spatio-temporal relationship between temperature and precipitation in the Caspian Sea. 
\cite{tsiotas2025} propose semi-parametric quantile trend models to analyze the dynamics of maximum and minimum temperatures in various regions of Greece.
However, these methodologies lack a spatial modeling structure, which is essential for capturing and quantifying the spatial dependencies and uncertainties inherent in most climatic data. 

Bayesian Gaussian process (GP) modeling is a gold standard for analyzing spatial and spatio-temporal data due to its ability to provide accurate predictions with rigorous uncertainty quantification \citep{banerjee2025}. In the context of spatial QR, \cite{castillo2023} develop a spatial quantile autoregressive model relying on four distinct GPs. However, the framework does not account for atmospheric covariates and requires a computationally intensive procedure to derive marginal quantiles. \cite{rosci2026} apply a Bayesian spatial QR model for high air-pollution concentrations using ERA5 reanalysis data as spatio-temporal predictors. However, spatial dependence is captured solely through a spatially varying intercept.

The main contribution of this manuscript is twofold. From an applied perspective, it provides a spatio-temporal characterization of the evolution of summer temperature quantiles over peninsular Spain, disentangling the variability associated with atmospheric circulation (AC) from its long-term temporal evolution. Specifically, daily maximum temperatures from meteorological stations are combined with ERA5 geopotential heights at three tropospheric levels, allowing atmospheric dynamics to be accounted for in the analysis.
Methodologically, this work introduces a Bayesian spatially varying coefficients model for perfect-prognosis statistical downscaling of temperature quantiles \citep{maraun2018}. By allowing the temperature response to a given atmospheric configuration to vary spatially and incorporating spatial-month-dependent long-term trends (LT), the proposed model disentangles the dynamical and thermodynamic drivers of regional climate change. Specifically, it separates changes in temperature associated with shifts in AC patterns from the background local warming trend.
Implemented via a Metropolis-within-Gibbs MCMC algorithm, the proposed framework provides spatial interpolation of marginal quantiles with complete measures of uncertainty, enabling the analysis to go beyond identifying mean warming, providing insight into how the interaction between AC and LT contributes to the spatial, temporal, and quantile heterogeneity of summer warming.

The proposed methodology is applied to the $0.50$ and $0.95$ quantiles of daily maximum temperatures observed from 40 meteorological stations located across peninsular Spain, covering from June to August from 1960 to 2023. This region spans about $492{,}000$ km$^2$ and features strong spatial variability in topography and climate. Comparing the two quantiles allows the assessment of whether changes in median and high temperatures differ in magnitude across space and time.

The format of the manuscript is as follows. Section~\ref{sec:data} presents the data and an exploratory data analysis. Section~\ref{sec:methodology} introduces the QR formulation, the proposed spatially varying quantile downscaling models, details of the model fitting, and goodness-of-fit metrics. Section~\ref{sec:analysis_data} shows the results of the application to the data described. Section~\ref{sec:conclusions} summarizes the main contributions and suggests future work to be explored. 

\section{Data and exploratory analysis}\label{sec:data}

This section introduces the daily maximum temperature data used as the response variable, the ERA5 geopotential fields used as predictors, and an exploratory analysis describing their relationship along with LT and variations across temperature quantiles.

\subsection{Daily maximum temperature data}

\begin{figure}[tb]
    \centering
    \includegraphics[width=\linewidth]{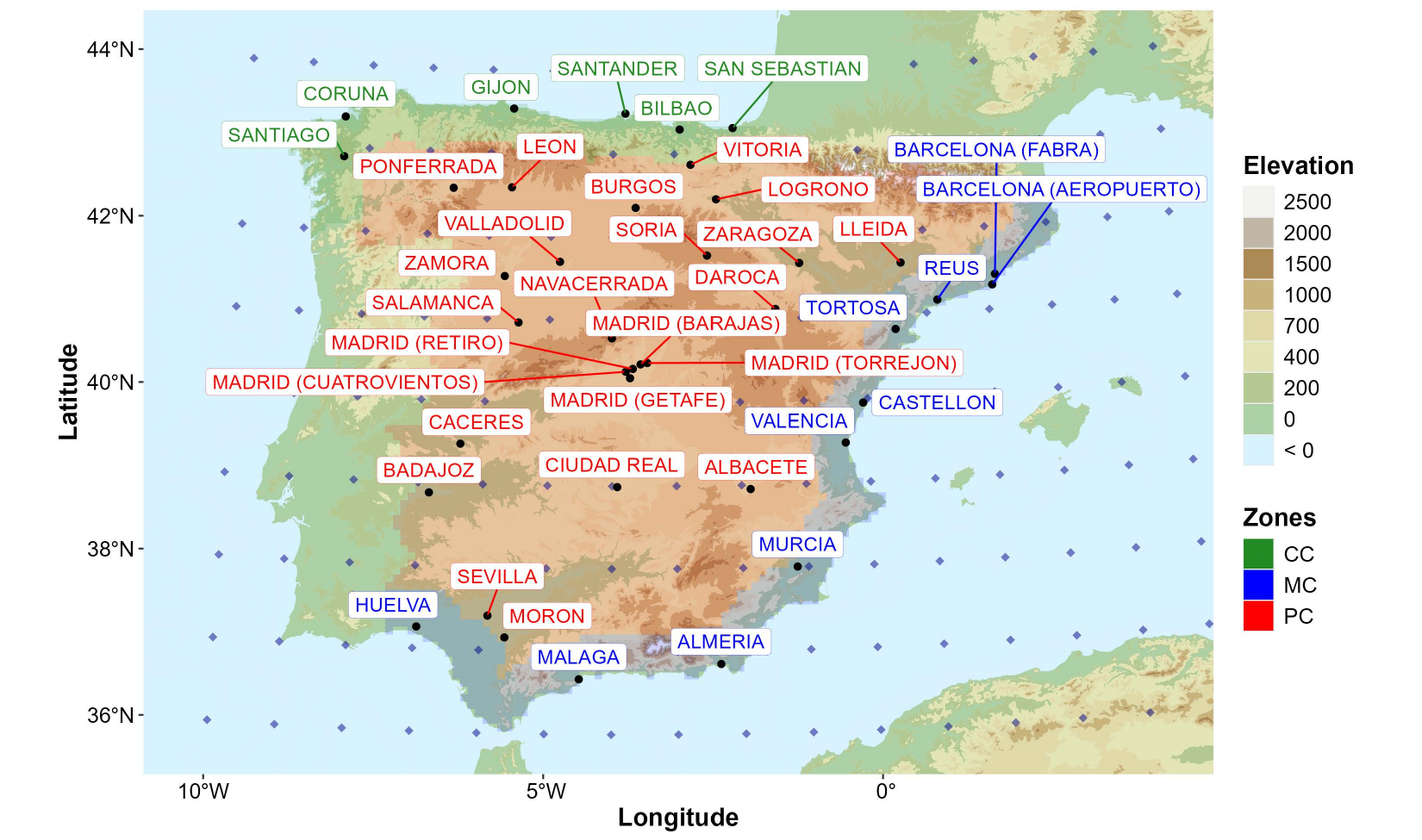}
    \caption{Map of meteorological stations in peninsular Spain, divided into climatic zones, overlaid on the ERA5 geopotential grid points.}
    \label{fig:mapas_est}
\end{figure}

The methodology presented here focuses on the analysis of daily maximum temperature ($^\circ$C) from $n=40$ meteorological stations across peninsular Spain. The data were obtained from the European Climate Assessment \& Dataset \citep[ECA\&D;][]{klein2002} and cover the period 1960--2023 at a daily resolution. We restrict the analysis to the summer months of June, July, and August (JJA), corresponding to 92 days per year over the 64-year study period. This choice is motivated by the fact that this period presents the highest observed temperatures and subsequent risk of extreme heat. The stations are distributed across peninsular Spain to provide adequate spatial coverage and were selected from the complete ECA\&D network based on their spatial coverage and temporal availability \citep{castillo2026}. Additionally, each site is characterized by geographical covariates including latitude, longitude, elevation (m), and minimum distance to the coast (km). The latter is used to classify the region into three climatic zones, Cantabrian coast (CC), Mediterranean coast (MC), and Peninsular Center (PC), with the coastal zones defined by a distance of less than $50$~km to the sea (see Figure~\ref{fig:mapas_est}).

As is common in observational climate records, some station series contain missing observations. Overall, missing values represent approximately $0.07\%$ of the complete dataset: 21 stations contain complete data, whereas for the remaining $19$ stations, the maximum number of missing values found is $44$, which accounts for $0.74\%$ of its individual time series. 
Rather than performing a separate preprocessing or imputation step, missing temperature observations are treated as latent quantities within the proposed Bayesian model and inferred jointly with the model parameters.

The region of study $D$ is peninsular Spain, with a spatial extent of $492{,}175$ km$^2$. 
In Spain, several mountain ranges can be found, including the Pyrenees in the northeast, the Inner Plateau in the center, and the Sierra Nevada along the southeastern coast. The country also has an extensive coastline, bordered by the Atlantic Ocean to the north and west, and the Mediterranean Sea to the south and east.
Spain is characterized by a wide variety of climates \citep{ign_clima_espana}. In the north, an oceanic climate predominates, while across the rest of the peninsula, climates range from warm semi-arid to Mediterranean oceanic, with the Mediterranean climate being dominant in the southeast.

\subsection{Atmospheric circulation predictors}

In addition to the temperature observations, we use AC predictors derived from the ERA5 reanalysis dataset \citep{hersbach2020,hersbach2023}. Specifically, we consider daily geopotentials ($\mathrm{m^2\,s^{-2}}$) at the $300$, $500$, and $700$ hPa pressure levels, available over the same period as the temperature data, recorded at 12:00 Coordinated Universal Time (UTC). The geopotential fields are provided on a regular $1^\circ \times 1^\circ$ grid covering peninsular Spain (see Figure~\ref{fig:mapas_est}). 

We use these three pressure levels to characterize the synoptic-scale atmospheric state at different levels of the troposphere while maintaining a parsimonious predictor set \citep[see, e.g.,][]{frauenfeld2003}. For each temperature station, the corresponding geopotential predictors are obtained from the ERA5 grid point closest to the station,  
providing a direct and contemporaneous representation of the AC at the observation location \citep{berrocal2010}. More complex predictors could incorporate nonlinear effects, interactions, temporal lags, information from multiple surrounding grid cells, or dimension-reduction approaches such as principal component analysis \citep{pryor2019}. We instead adopt this simpler formulation as a parsimonious and interpretable representation, avoiding unnecessary model complexity \citep{zorita1999}.

In this work, we focus on daily geopotential anomalies, since a statistical downscaling method based on anomalies tends to produce a more realistic picture of the local climate \citep{benestad2008}. For each geopotential level at every location, we define these anomalies as the residuals of a least squares linear regression model fitted to the standardized daily geopotentials at 12:00~UTC. The models incorporate an intercept and a first-order harmonic term to account for the seasonal cycle, and are fitted over the 1981--2010 reference period following World Meteorological Organization standards \citep{wmo2017}.

\subsection{Exploratory analysis}\label{sec:desc_an}

This section presents an exploratory analysis to characterize the spatio-temporal dynamics of daily maximum temperatures and ERA5 geopotential fields. First, we decompose both signals into long-term temporal evolution and seasonal components, comparing their standardized seasonal cycles across climatic zones. Second, we assess the differences in quantile warming across both 32-year periods. Third, we evaluate global and month-specific quantile trends, examining their relationship with distance to the coast and elevation. Finally, we map the spatial distribution of the $0.50$ and $0.95$ empirical quantiles and evaluate the effects of LT and AC via local frequentist QR models, motivating the modeling choices in subsequent sections. 

Figures~S1 and S2 of the Supplementary material (hereinafter, `S' denotes Supplementary material) show clear long-term temporal evolutions and seasonal components in the summer period for both daily maximum temperatures and ERA5 geopotentials, revealing similar behaviors between variables. However, although slight differences in their estimated seasonal components may be found (see Figure~S3), working with daily geopotential anomalies effectively removes this seasonal component, leaving a signal focused on synoptic variability. 

\begin{figure}
    \centering
    \includegraphics[width=0.85\linewidth]{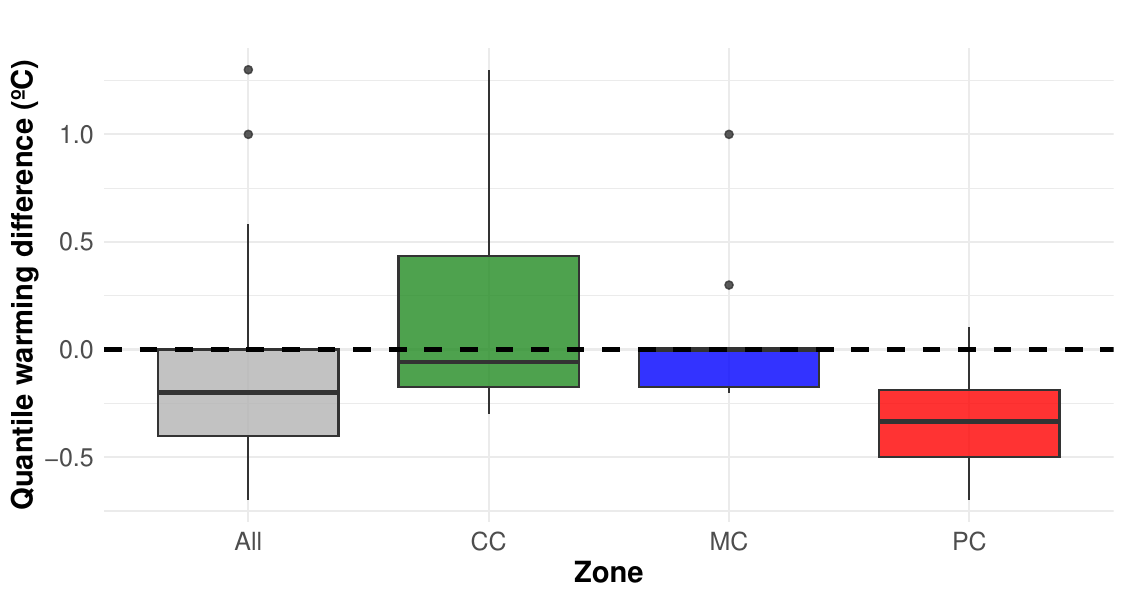}
    \caption{Quantile warming difference between the $0.95$ and the $0.50$ quantiles between the 2023--1992 and 1960--1991 JJA periods for each climatic zone.}
    \label{fig:bp_warming}
\end{figure}

To motivate the fitting of separate models for each quantile, we analyze the evolution of quantiles across the study period. For each available station, we first compute the empirical $0.95$ and $0.50$ quantiles for the $1992$--$2023$ and $1960$--$1991$ 32-year periods. Then, quantile warming is defined as the difference between the previous values. Later, we define the quantile warming difference as the subtraction between the $0.95$ and $0.50$ quantile warmings. Figure \ref{fig:bp_warming} displays the distribution of this quantile warming difference across stations in each climatic zone. Values greater than zero indicate that the $0.95$ quantile has warmed more rapidly than the $0.50$ quantile. Clear differences in both central tendency and variability are visible across the different climatic zones. In the PC, most of the values are negative, indicating greater warming in the $0.50$ quantile than in the $0.95$ quantile. On the MC, the warming of the quantiles seems similar, with the warming of the median slightly higher. Lastly, on the CC, there is high variability in the differences, and mainly positive, indicating greater warming of the high quantile.

Thermal trends can generally be decomposed into a dynamic component driven by synoptic AC and a residual linear trend associated with background warming. In a model where AC is accounted for, the explicit trend term captures this residual background warming. Section~S1.1 presents an exploratory analysis of quantile trends for representative stations in each climatic zone. Results reveal marked differences across quantiles, zones, and months. Because these warming rates vary across the summer season, the empirical patterns motivate the inclusion of a month-dependent LT. Additionally, this supplementary section evaluates the spatial relationship between LT magnitudes and distance-to-coast across stations, showing a linear relationship (especially for the $0.50$ quantile) with unexplained variability unrelated to the station elevation.

Then, we analyze the relationship between the empirical $0.95$ and $0.50$ quantiles in the whole study period and geographical variables. Figure~S8 reflects a clear spatial dependence of the empirical quantiles with respect to the distance to the coast and elevation at which the meteorological stations are located. Coastal and high-altitude stations present lower empirical quantiles compared to those stations located in the PC.

\begin{figure}
    \centering
    \includegraphics[width=\linewidth]{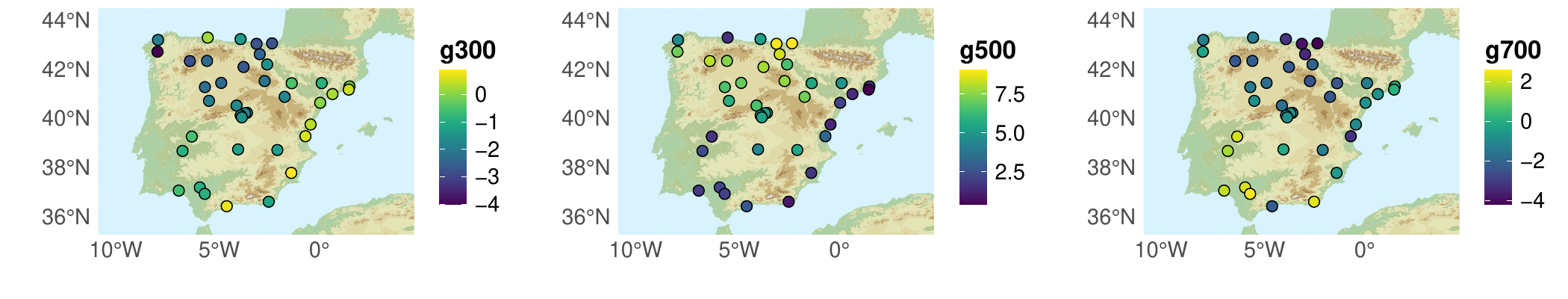}
    \caption{Estimated standardized geopotential anomalies coefficients of the local $0.95$ QR model across stations.}
    \label{fig:coef_geop_g300}
\end{figure}

Lastly, we evaluate the effect of the AC and LT on the $0.50$ and $0.95$ daily maximum temperature quantiles at each station through local QR models with the \texttt{quantreg} R package \citep{koenker2016}. The fitted models include an intercept, a first-order harmonic term, month-dependent LT, and geopotential anomalies, yielding a total of 9 regression coefficients. Figure \ref{fig:coef_geop_g300} shows the estimated coefficients accompanying the standardized geopotential anomalies for every station for the $0.95$ quantile. A clear spatial dependence is evident, motivating the incorporation of spatially varying coefficients dependent on the distance-to-coast. Standardized seasonal components and LT exhibit similar results (see Figure~S9), as do the estimated coefficients for the $0.50$ QR models (see Figure~S10).

\section{Methodology} \label{sec:methodology}

This section describes the statistical methodology used to develop and evaluate the proposed Bayesian spatially varying quantile downscaling framework. We first introduce the Bayesian formulation of QR based on the AL distribution. We then present the spatially varying quantile downscaling models considered in this study, followed by the model fitting procedure, model-based spatial interpolation of quantiles, and the metrics used for model selection and goodness-of-fit assessment.

\subsection{Bayesian quantile regression via the AL distribution}

We first introduce the Bayesian formulation of QR based on the AL distribution proposed by \cite{yu2001}, which provides the basis for the spatio-temporal QR model developed in the following sections. Let $Y_i$ denote a response variable and $\mathbf{x}_i$ a vector of covariates. For a given quantile level $\tau \in (0,1)$, QR models the conditional $\tau$-quantile of $Y_i$ given $\mathbf{x}_i$ as
\begin{equation*}
Y_i = Q_{Y_i \mid \mathbf{x}_i}(\tau) + \epsilon_i = \mathbf{x}_i^\top \bm{\beta} + \epsilon_i,
\end{equation*}
where $P(\epsilon_i \leq 0) = \tau$. Thus, $Q_{Y_i \mid \mathbf{x}_i}(\tau) = \mathbf{x}_i^\top\bm{\beta}$ represents the conditional $\tau$-quantile of $Y_i$ given $\mathbf{x}_i$. In the Bayesian formulation, the error term is assumed to follow an AL distribution, denoted by $\epsilon_i \sim \mathrm{AL}(\mu,\sigma,\tau)$, where $\mu$ is a location parameter, $\sigma > 0$ is a scale parameter, and $\tau \in (0,1)$ is an asymmetry parameter. For a fixed quantile level $\tau$ corresponding to the quantile of interest, the location parameter $\mu$, which is the $\tau$-quantile of the distribution, is set to $\mu=0$ in the QR model.

The connection between the AL distribution and QR is given by the quantile loss function $\delta_\tau(u) = u\left[\tau-\mathbf{1}(u<0)\right]$ with $\mathbf{1}(\cdot)$ the indicator function. The conditional density of $Y_i$ is then
\begin{equation*}
f(Y_i \mid Q_{Y_i \mid \mathbf{x}_i}(\tau),\sigma,\tau)
=
\frac{\tau(1-\tau)}{\sigma}
\exp\left\{
-\frac{\delta_\tau\left(Y_i-Q_{Y_i \mid \mathbf{x}_i}(\tau)\right)}
{\sigma}
\right\}.
\end{equation*}
Hence, the AL likelihood is directly related to the quantile loss function, with $Q_{Y_i \mid \mathbf{x}_i}(\tau)$ determining the conditional $\tau$-quantile.

For Bayesian computation, we use the normal-exponential mixture representation of the AL distribution introduced by \cite{kozumi2011}. Let $Z_i \sim N(0,1)$ and $\xi_i \sim \sigma \, Exp(1)$ be independent. Then,
\begin{equation}\label{eq:normal_exp_rep}
\epsilon_i
=
\theta \xi_i
+ \omega \sqrt{
\sigma \xi_i
}\,Z_i, \quad \text{with} \quad \theta = \frac{(1-2\tau)}
{\tau(1-\tau)}, \quad \omega = \sqrt{
\frac{2 }
{\tau(1-\tau)}
}.
\end{equation}
Consequently, conditional on the latent exponential variable $\xi_i$, the model has the Gaussian representation
\begin{equation*}
Y_i \mid Q_{Y_i \mid \mathbf{x}_i}(\tau),\sigma,\tau,\xi_i
\sim
N\left(
Q_{Y_i \mid \mathbf{x}_i}(\tau)
+
\theta \xi_i,
\omega^2 \sigma \xi_i
\right).
\end{equation*}
This representation facilitates Bayesian inference through Gibbs sampling by introducing the latent variables $\xi_i$.

\subsection{Spatially varying quantile downscaling models} \label{sec:model}

We consider three quantile downscaling models to assess the respective contributions of LT and large-scale AC to local summer temperature. All models include a spatial baseline and seasonal terms, while differing in their inclusion of LT, AC predictors, or both. We first compare alternative specifications for the spatial structure of the regression coefficients based on out-of-sample predictive performance. We then compare the three models in terms of in-sample goodness-of-fit, using the selected spatial specification (see Section~\ref{sec:model_selection}).

Let $Y_{t\ell}(\mathbf{s})$ denote the daily maximum temperature observed at location $\mathbf{s} \in D$, for year $t=1 \,(1960),\ldots,T = 64 \, (2023)$ and summer day $\ell=152 \, (\text{June 1st}),\ldots, L = 243 \, (\text{August 31st})$, where $D$ denotes the region of interest. Geopotential is available on a regular spatial grid. We denote by $G_{t\ell}^{h}(\mathbf{b})$ the geopotential at pressure level $h \in H = \{300,500,700\}$ for grid point $\mathbf{b} \in D$ on day $(t,\ell)$. Each location $\mathbf{s}$ is associated with its closest grid point $\mathbf{b}$, so that all locations with the same closest grid point are assigned the same geopotential value. Thus, the geopotential fields are discretized at the grid-point resolution, with each grid point providing a geopotential value for each pressure level and day.

Our formulation builds on the spatially varying coefficient approach to statistical downscaling of \cite{berrocal2010}, extending it to a QR framework to characterize the conditional temperature distribution beyond its mean. For a given quantile level $\tau \in (0,1)$, the most general model considered is
\begin{equation} \label{eq:model}
\begin{aligned}
Y_{t\ell}(\mathbf{s})
={}&
\underbrace{\tilde{\beta}_0(\mathbf{s})
+
\tilde{\beta}_1(\mathbf{s})
\sin\left(2\pi\ell / 365\right)
+
\tilde{\beta}_2(\mathbf{s})
\cos\left(2\pi\ell / 365\right)}
_{\text{baseline and seasonal terms}}
\\ & +
\underbrace{
\sum_{k=1}^{3}
\tilde{\gamma}_k(\mathbf{s})\,
\mathbf{1}\{m(l)=k\}\,t
}_{\text{LT}}
+
\underbrace{
\sum_{h \in {H}}
\tilde{\alpha}_h(\mathbf{s})\,
G_{t\ell}^{h}(\mathbf{b})
}_{\text{AC}}
+
\epsilon_{t\ell}(\mathbf{s}),
\end{aligned}
\end{equation}
where $m(\ell)$ denotes the month corresponding to summer day $\ell$, with $m(\ell)=1,2,3$ corresponding to June, July, and August, respectively. Motivated by the descriptive analysis carried out in Section \ref{sec:desc_an}, the spatially varying coefficients are specified as
\begin{equation} \label{eq:svc}
\begin{aligned}
    &\tilde{\beta}_0(\mathbf{s}) = \beta_{00} + \beta_{01} \log(1 + \dist(\mathbf{s})) + \beta_{02} \elev(\mathbf{s}) + \beta_0(\mathbf{s}), & \\
    &\tilde{\beta}_j(\mathbf{s}) = \beta_{j0} + \beta_{j1} \log(1 + \dist(\mathbf{s})) + \beta_j(\mathbf{s}), & j = 1,2, \\
    &\tilde{\gamma}_k(\mathbf{s}) = \gamma_{k0} + \gamma_{k1} \log(1 + \dist(\mathbf{s})) + \gamma_k(\mathbf{s}), & k = 1,2,3, \\
    &\tilde{\alpha}_h(\mathbf{s}) = \alpha_{h0} + \alpha_{h1} \log(1 + \dist(\mathbf{s})) + \alpha_h(\mathbf{s}), & h \in H,
\end{aligned}
\end{equation}
and $\epsilon_{t\ell}(\mathbf{s}) \sim \text{i.i.d. } \mathrm{AL}(0,\sigma,\tau)$ for QR. Here, $\dist(\mathbf{s})$ and $\elev(\mathbf{s})$ denote the distance to the coast in km and elevation in meters at location $\mathbf{s}$, respectively. The coefficients $\beta_{j0}$, $\gamma_{k0}$, and $\alpha_{h0}$ represent the corresponding global effects, while the terms involving $\log(1+\dist(\mathbf{s}))$ account for systematic spatial variation with distance to the coast. For the spatial intercept, elevation is additionally included to account for systematic variation associated with topography. The terms $\beta_j(\mathbf{s})$, $\gamma_k(\mathbf{s})$, and $\alpha_h(\mathbf{s})$ represent local spatial adjustments that capture the remaining spatially structured variation in the corresponding coefficients.

As previously stated, to assess the relative contributions of LT and AC, we consider three model specifications. Model $M_1$ includes the spatial baseline, seasonal terms, and month-specific LT; model $M_2$ includes the spatial baseline, seasonal terms, and AC predictors; and model $M_3$ is the full model in \eqref{eq:model}, including both LT and AC predictors.

\subsubsection{Model specifics}\label{sec:model_specifics}

The local spatial adjustments in \eqref{eq:svc} are modeled as spatial GPs. Specifically, for each of the nine coefficient surfaces, we assume an independent zero-mean GP with an exponential covariance function,
\begin{equation*}
\begin{aligned}
\beta_j(\mathbf{s}) &\sim GP\left(0, \sigma_{\beta_j}^2
\exp\left\{-\phi_{\beta_j}\|\mathbf{s}-\mathbf{s}'\|\right\}\right), & j = 0,1,2,\\
\gamma_k(\mathbf{s}) &\sim GP\left(0, \sigma_{\gamma_k}^2
\exp\left\{-\phi_{\gamma_k}\|\mathbf{s}-\mathbf{s}'\|\right\}\right), & k = 1,2,3,\\
\alpha_h(\mathbf{s}) &\sim GP\left(0, \sigma_{\alpha_h}^2
\exp\left\{-\phi_{\alpha_h}\|\mathbf{s}-\mathbf{s}'\|\right\}\right), & h \in H.
\end{aligned}
\end{equation*}
Here, $\|\mathbf{s}-\mathbf{s}'\|$ denotes the Euclidean distance between locations $\mathbf{s}, \mathbf{s}' \in D$; $\sigma_{\beta_j}^2$, $\sigma_{\gamma_k}^2$, and $\sigma_{\alpha_h}^2$, denote the marginal variances of the respective spatial processes; while $\phi_{\beta_j}$, $\phi_{\gamma_k}$, and $\phi_{\alpha_h}$, control their spatial decay. This specification induces spatial dependence between nearby locations while allowing the coefficient surfaces to vary continuously over the study region. Consequently, the spatial processes enable interpolation of the coefficient surfaces and, through them, of the corresponding conditional temperature quantiles at unobserved locations in $D$.

Whenever possible, we use independent and conditionally conjugate priors to facilitate Gibbs sampling. For the non-spatial regression coefficients in \eqref{eq:svc}, we assign independent diffuse normal priors with zero mean and standard deviation $100$. The residual scale parameter is assigned an inverse-gamma prior, with shape-rate parameterization $\sigma \sim IG(0.1,0.1)$.
The hyperparameters in the spatial processes are assigned inverse-gamma priors, $\sigma_{\beta_j}^2, \sigma_{\gamma_k}^2, \sigma_{\alpha_h}^2 \sim IG(0.1, 0.1)$, and gamma priors $\phi_{\beta_j}, \phi_{\gamma_k}, \phi_{\alpha_h} \sim G(38, 7400)$. In the case of the decay hyperparameters, we have considered more informative priors to address the well-known identifiability issues associated with them \citep{zhang2004}. This specification yields a prior for the effective range with a prior mean of $600$ km, corresponding to approximately two thirds of the maximum distance between stations, and a standard deviation of $100$ km; i.e. it places $95\%$ of the probability mass between $435$ and $825$ km.

\subsection{MCMC implementation and spatial interpolation}

Model fitting and inference are conducted within a fully Bayesian hierarchical framework using MCMC methods implemented in the \texttt{spTReg} R package \citep{castillo2026spTReg}, allowing for a comprehensive quantification of uncertainty.

\subsubsection{MCMC implementation}

Here, we describe the full conditional distributions required for Metropolis-within-Gibbs sampling. Following the structure in \eqref{eq:model}, any spatially varying QR model at a given quantile level $\tau \in (0, 1)$ can be written as
\begin{equation} \label{eq:general_model_MCMC}
    Y_{t\ell}(\mathbf{s}) = \sum_{m = 1}^{r} x_{t\ell, m}(\mathbf{s}) \beta_m(\mathbf{s}) + \epsilon_{t\ell}(\mathbf{s}), \quad \epsilon_{t\ell}(\mathbf{s}) \sim \text{i.i.d. } AL(0, \sigma, \tau),
\end{equation}
where $r$ is the total number of spatio-temporal covariates, $x_{t\ell, m}(\bs)$ is the $m$-th covariate value in day $\ell$ within year $t$ at location $\bs$, and the spatially-varying coefficients are modeled as $\beta_m(\bs) \sim GP(\mathbf{z}_m(\bs)^\top\bm\gamma_m, \sigma^2_{m} R(\bs, \bs'; \phi_{m}))$. We use hierarchical centering for the spatial GPs \citep{gelfand1995}, each mean function is modeled via a spatial linear predictor with spatial covariates $\mathbf{z}_m(\bs)$ and coefficients vector $\bm\gamma_m$, and the correlation function is $R(\bs, \bs'; \phi_{m}) = \exp\left\{-\phi_{m}\|\mathbf{s}-\mathbf{s}'\|\right\}$.  

For the set of monitoring stations, $S = \{\bs_1, \dots, \bs_n\}$, under the GP model, for $m = 1, \ldots, r$, we have
\begin{align*}
    \bm\beta_m = (\beta_m(\bs_1), \dots, \beta_m(\bs_n))^\top \sim N_n(\mathbf{Z}_m\bm\gamma_m, \sigma^2_{m}\mathbf{R}_{\phi_{m}}),
\end{align*}
where $\mathbf{Z}_m = (\mathbf{z}_m(\bs_1), \dots, \mathbf{z}_m(\bs_n))^\top$ corresponds to the $n\times p_m$ design matrix containing an intercept and $p_m - 1$ spatial variables, and $\bm\gamma_m$ denotes the $p_m$-dimensional vector of regression coefficients. The correlation matrix $\mathbf{R}_{\phi_m}$ is obtained by evaluating the correlation function $R(\bs, \bs'; \phi_{m})$ at the monitoring stations.

We define the following vectors and diagonal matrix: $\mathbf{Y}_{t\ell} = (Y_{t\ell}(\bs_1), \ldots, Y_{t\ell}(\bs_n))^\top$, $\bm{\epsilon}_{t\ell} = (\epsilon_{t\ell}(\bs_1), \ldots, \epsilon_{t\ell}(\bs_n))^\top$, and $\mathbf{X}_{t\ell, m} = \operatorname{diag}(x_{t\ell, m}(\bs_1), \ldots, x_{t\ell, m}(\bs_n))$. Then, model \eqref{eq:general_model_MCMC} in matrix  notation is
\begin{equation*}
    \mathbf{Y}_{t\ell} = \sum_{m = 1}^r \mathbf{X}_{t\ell, m} \bm\beta_m + \bm\epsilon_{t\ell}.
\end{equation*}

The latent exponential variables $\xi_{t\ell}(\bs_i)$, collectively denoted by $\bm\xi$, are introduced through the normal-exponential representation of the AL distribution in \eqref{eq:normal_exp_rep} to facilitate Gibbs sampling. Furthermore, let $\mathbf{Y}$ denote the vector collecting all response observations.

\paragraph{Full conditional distribution of $\xi_{t\ell}(\bs_i)$} For $t=1,\ldots,T$, $\ell=1,\ldots,L$, and $i = 1,\ldots,n$, we sample values of $\xi_{t\ell}(\mathbf{s}_i)$ from its generalized inverse Gaussian full conditional distribution,
\begin{equation*}
    \xi_{t\ell}(\mathbf{s}_i) \mid \{\bm\beta_m\}_{m = 1}^r, \sigma, \mathbf{Y} \sim 
    GIG\left( 
    \frac{1}{\sigma} \left( 2 + \frac{\theta^{2}}{\omega^{2}} \right),
    \frac{Y_{t\ell}^{*}(\bs_i)^2}{\omega^2 \sigma},
    \frac{1}{2}
    \right),
\end{equation*}
where $Y_{t\ell}^{*}(\bs_i) = Y_{t\ell}(\mathbf{s}_i) - \sum_{m = 1}^{r} x_{t\ell, m}(\mathbf{s}_i) \beta_m(\mathbf{s}_i)$ denotes the adjusted response after removing the contribution of all regressors.

The employed GIG distribution has a density function proportional to 
\begin{equation*}
    f(x\mid \psi, \chi,\lambda)\propto x^{\lambda-1} \exp\left\{-\frac{1}{2}(\chi/x+\psi x)\right\}.
\end{equation*}

\paragraph{Full conditional distribution of $1 / \sigma$} Given a gamma prior distribution $G(a, b)$ for $1 / \sigma$, we sample values of $1/\sigma$ from its gamma full conditional distribution, 
\begin{equation*}
    \frac{1}{\sigma} \mid \{\bm\beta_m\}_{m = 1}^r, \bm\xi, \mathbf{Y} \sim
    G \left( a + \frac{3}{2}TLn, b + \sum_{t=1}^{T} \sum_{\ell=1}^{L} \sum_{i=1}^{n} \xi_{t\ell}(\mathbf{s}_i) + \frac{1}{2\omega^2} \sum_{t=1}^{T} \sum_{\ell=1}^{L} \sum_{i=1}^{n} \frac{Y_{t\ell}^{*}(\bs_i)^2}{\xi_{t\ell}(\mathbf{s}_i)} \right),
\end{equation*}
where $Y_{t\ell}^{*}(\bs_i) = Y_{t\ell}(\mathbf{s}_i) - \sum_{m = 1}^{r} x_{t\ell, m}(\mathbf{s}_i) \beta_m(\mathbf{s}_i) - \theta\xi_{t\ell}(\mathbf{s}_i)$ denotes the adjusted response after removing the contribution of the latent variable and all regressors.

\paragraph{Full conditional distribution of \texorpdfstring{$\bm{\beta}_m$}{beta	extunderscore m}} For $m = 1, \ldots, r$, we sample values of $\bm\beta_m$ from its multivariate normal full conditional distribution,
\begin{align*}
    \bm{\beta}_m \mid \bm\gamma_m,  \sigma^2_{m}, \phi_{m}, \bm\xi, \mathbf{Y} &\sim N_n(\mathbf{m}_{\bm{\beta}_m}, \mathbf{C}_{\bm{\beta}_m}); \\
    \mathbf{C}_{\bm{\beta}_m}^{-1} & = \sum_{t = 1}^T \sum_{\ell = 1}^L \mathbf{X}_{t\ell, m}^\top \bm\Omega_{t\ell}^{-1} \mathbf{X}_{t\ell, m} + \frac{1}{\sigma_m^2}\mathbf{R}_{\phi_{m}} ^{-1},\\
    \mathbf{m}_{\bm{\beta}_m} & =  \mathbf{C}_{\bm{\beta}_m} \left(\sum_{t = 1}^T \sum_{\ell = 1}^L \mathbf{X}_{t\ell, m}^\top \bm\Omega_{t\ell}^{-1} \mathbf{Y}_{t\ell, m}^{*}  + \frac{1}{\sigma_m^2}\mathbf{R}_{\phi_{m}} ^{-1}\mathbf{Z}_m\bm\gamma_m\right),
\end{align*} 
where $\mathbf{Y}_{t\ell, m}^{*} = (Y_{t\ell, m}^{*}(\bs_1), \ldots, Y_{t\ell, m}^{*}(\bs_n))^\top$ 
with $Y_{t\ell, m}^{*}(\bs_i) = Y_{t\ell}(\bs_i) - \sum_{j \neq m}x_{t\ell, j}(\bs_i)\beta_j(\bs_i) - \theta\xi_{t\ell}(\bs_i)$ denotes the adjusted response after removing the contribution of all regressors except the $m$-th one and the latent variable, and $\bm\Omega_{t\ell} = \omega^2 \sigma \operatorname{diag}( \xi_{t\ell}(\bs_1), \ldots, \xi_{t\ell}(\bs_n))$ denotes the diagonal conditional covariance matrix of the observations across the $n$ monitoring stations, given the latent exponential variables.

\paragraph{Full conditional distribution of the Gaussian process hyperparameters}
For the spatially varying coefficients, we also need to sample the hyperparameters corresponding to the mean regression coefficients, the variance, and the decay parameter. So, for $m = 1, \ldots, r$, we proceed as follows.

Given a normal prior distribution $N_{p_m} \left(\mathbf{a}, \mathbf{B} \right)$ for $\bm{\gamma}_m$, we sample values of $\bm{\gamma}_m$ from its multivariate normal full posterior distribution,
\begin{align*}
    \bm\gamma_m \mid \bm\beta_m, \sigma^2_{m}, \phi_{m} &\sim N_{p_m} (\mathbf{m}_{\bm\gamma_m}, \mathbf{C}_{\bm\gamma_m}); \\ 
    \mathbf{C}_{\bm\gamma_m}^{-1} &= \frac{1}{\sigma_{m}^2} \mathbf{Z}_m^\top \mathbf{R}^{-1}_{\phi_{m}} \mathbf{Z}_m + \mathbf{B}^{-1}, \\
    \mathbf{m}_{\bm\gamma_m} &= \mathbf{C}_{\bm\gamma_m} \left( \frac{1}{\sigma_{m}^2} \mathbf{Z}_m^\top \mathbf{R}^{-1}_{\phi_{m}} \bm{\beta}_m + \mathbf{B}^{-1} \mathbf{a} \right).
\end{align*}

Given a gamma prior distribution $G(a, b)$ for $1 / \sigma^2_{m}$, we sample values of $1 / \sigma^2_{m}$ from its gamma full conditional distribution,
\begin{equation*}
    \frac{1}{\sigma_{m}^2} \mid \bm\beta_m, \, \bm\gamma_m, \, \phi_{m} \sim G\left( a + \frac{n}{2}, b + \frac{1}{2} (\bm{\beta}_m -  \mathbf{Z}_m \bm{\gamma}_m)^\top \mathbf{R}^{-1}_{\phi_{m}}(\bm{\beta}_m - \mathbf{Z}_m \bm{\gamma}_m)
    \right).
\end{equation*}

Given a gamma prior distribution $G(a,b)$ for $\phi_m$, its full conditional distribution is not of a standard form. Therefore, we update $\phi_m$ on the logarithmic scale using an adaptive random-walk Metropolis step \citep{roberts2009}. The proposal follows a Gaussian random walk with adaptively tuned variance, targeting an acceptance rate of 33\%. The target kernel is given by
\begin{align*}
    &[ \log\phi_{m} \mid \bm\beta_m, \bm\gamma_m, \sigma^2_{m} ]
    \\ &\quad \propto 
    \lvert \mathbf{R}_{\phi_{m}} \rvert^{-1/2} \cdot 
    \exp\left\{ -\frac{1}{2\sigma_{m}^2} (\bm{\beta}_m - \mathbf{Z}_m \bm{\gamma}_m)^\top \mathbf{R}_{\phi_{m}}^{-1} (\bm{\beta}_m - \mathbf{Z}_m \bm{\gamma}_m) + a \log\phi_{m} - b \phi_{m} \right\}.
\end{align*}

\subsubsection{Spatial interpolation}\label{sec:spatial_interpolation}

The spatial GPs provide a direct approach for interpolating the regression coefficients at unobserved locations. Let $\mathbf{s}_0 \in D$ denote a new location. For each MCMC iteration, we obtain a realization of the spatial GPs at $\mathbf{s}_0$ from their conditional Gaussian distribution given the observed locations, this procedure is known as Bayesian kriging \citep{banerjee2025}. These realizations are then combined with the corresponding global coefficients and spatial covariates to obtain posterior samples of the spatially varying coefficients in \eqref{eq:svc}.

For a given year $t$, and summer day $\ell$, each posterior realization of these coefficients is used to evaluate the corresponding conditional temperature quantile given the vector $\mathbf{G}_{t\ell}(\mathbf{b}_0)$, with $\mathbf{b}_0$ the closest grid point to $\bs_0$:
\begin{align*} \label{eq:quantile_interpolation}
Q_{Y_{t\ell}(\mathbf{s}_0) \mid \mathbf{G}_{t\ell}(\mathbf{b}_0)}(\tau)
={}&
\tilde{\beta}_0(\mathbf{s}_0)
+
\tilde{\beta}_1(\mathbf{s}_0)
\sin\left(2\pi\ell / 365\right)
+
\tilde{\beta}_2(\mathbf{s}_0)
\cos\left(2\pi\ell / 365\right)
\\
&+
\sum_{k=1}^{3}
\tilde{\gamma}_k(\mathbf{s}_0)\,
\mathbf{1}\{m(\ell)=k\}\,t
+
\sum_{h \in H}
\tilde{\alpha}_h(\mathbf{s}_0)\,
G_{t\ell}^{h}(\mathbf{b}_0).
\end{align*}
The resulting samples characterize the posterior distribution of the conditional quantile surface and allow posterior means and credible intervals (CIs) to be obtained at any location within the study region. Thus, uncertainty in the spatial coefficient processes is propagated directly to the interpolated temperature quantiles.

\subsection{Model selection and goodness of fit}
\label{sec:model_selection}

After fitting the proposed QR models, their performance is evaluated from two distinct perspectives: predictive validity and explanatory power. To avoid overfitting during model selection, we employ a cross-validation (CV) scheme and evaluate the metrics out-of-sample. Conversely, to measure the in-sample goodness of fit, we compute the metrics across the entire training dataset. Specifically, we focus on two metrics: the $R^1(\tau)$ coefficient, based on \cite{koenker1999}, and the empirical exceedance probability $p(\tau)$, adapted from \cite{castillo2023}.

\subsubsection{The R1 quantile coefficient}

The $R^1(\tau)$ metric serves as the direct QR analogue to the traditional $R^2$ coefficient commonly used in mean regression models. While $R^2$ measures the reduction in the sum of squared residuals, $R^1(\tau)$ quantifies the reduction in the sum of weighted absolute deviations (the quantile loss function) achieved by the full model relative to a baseline null model. The global $R^1(\tau)$ is formally defined as:
\begin{equation} \label{eq:R1}
    R^1(\tau) = 1 - \frac{V(\tau)}{V_{0}(\tau)},
\end{equation}
where $V(\tau)$ and $V_{0}(\tau)$ represent the total minimized quantile loss across all stations, years, and days for the full and null models, respectively. For the baseline null model, $V_{0}(\tau)$ is computed using the station-specific empirical $\tau$-th quantile.

When evaluated in-sample for goodness-of-fit assessment, $R^1(\tau)$ strictly ranges between 0 and 1, where higher values indicate a superior fit. However, when computed out-of-sample during cross-validation for model selection, $R^1(\tau)$ can take negative values if the predictive model performs worse than the baseline. In addition to this global metric, a localized spatial version denoted as $R^1(\tau; \mathbf{s}_i)$ can be computed by restricting the loss accumulation to a specific station $\mathbf{s}_i \in S$.

\begin{figure}[htbp]
    \centering
    \includegraphics[width=0.35\textwidth]{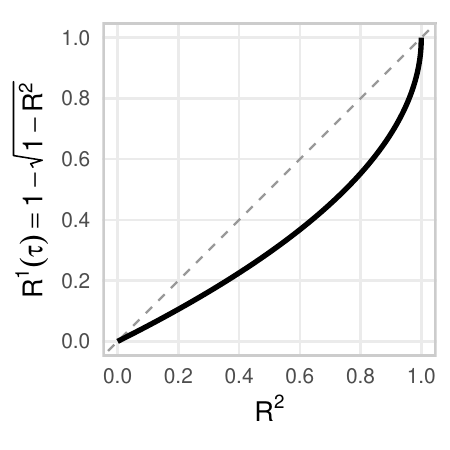}
    \caption{Theoretical relationship between the $R^2$ and the $R^1(\tau)$ coefficients under homoscedasticity and joint normality of the variables. The solid line represents the exact closed-form relation $R^1(\tau) = 1 - \sqrt{1 - R^2}$.}
    \label{fig:relationship_r1_r2}
\end{figure}

An intuition about the relationship between $R^1(\tau)$ and the traditional $R^2$ is often neglected. While $R^1(\tau)$ shares a 0-to-1 scale with the traditional $R^2$, their absolute magnitudes differ due to the distinct geometries of their loss functions. Under strict homoscedasticity and joint normality of the variables, the asymmetric weights of the quantile loss function cancel out. Consequently, $R^1(\tau)$ becomes invariant across all thresholds $\tau$ and collapses to the exact closed-form relationship $R^1(\tau) = 1 - \sqrt{1 - R^2}$ (e.g., an $R^2$ of $0.60$ yields an $R^1$ of approximately $0.37$). However, if the covariates follow a non-Gaussian distribution, the marginal distribution of the response becomes a complex convolution of densities. This breaks the mathematical cancellation in the denominator, causing $R^1(\tau)$ to lose its invariance and vary across different quantiles. Therefore, a lower absolute $R^1$ value does not necessarily indicate a poor fit, but rather reflects the linear penalty of absolute deviation metrics compared with the quadratic penalty (see Figure~\ref{fig:relationship_r1_r2}).

\subsubsection{Empirical exceedance probability}

The second metric, $p(\tau)$, evaluates whether the estimated conditional quantiles accurately capture the nominal probability threshold $\tau$ across the spatio-temporal domain. This metric tracks the empirical proportion of observed data points that fall below the model's predicted quantile. In particular, the global calibration measure $p(\tau)$ is obtained by averaging across all dimensions a local indicator function, which takes the value of 1 if the observed response on a given day, year, and station falls below its corresponding predicted conditional quantile, and 0 otherwise. 

Under a correctly specified and well-calibrated model, the expected value of this indicator must equal $\tau$. Consequently, a well-fitted model implies that the global $p(\tau)$ should closely approximate the nominal value of $\tau$. To identify potential localized patterns or structural miscalibrations, analogous versions can be computed by averaging across specific dimensions. These include variations over days $p_\ell(\tau)$, years $p_t(\tau)$, or individual spatial stations $p(\tau; \mathbf{s}_i)$.

\section{Analysis of the data}\label{sec:analysis_data}

This section presents the results of applying the proposed methodology of Section~\ref{sec:methodology} to the dataset introduced in~Section~\ref{sec:data}. It first discusses the specification for the spatial covariates incorporated inside the mean of the spatially varying coefficients. Subsequently, the relative contribution of the LT and the AC to explaining quantile variability is assessed. Finally, the section concludes with the inference results for the proposed model.

All models were fitted by scaling the covariates to have zero mean and unit variance to enhance MCMC mixing. Model fitting was performed by running two chains of the MCMC algorithm for each model, each starting with distinct initial values and running for $200{,}000$ iterations to collect samples from the joint posterior distribution. The first half of the samples were discarded as burn-in, and the remaining samples were thinned to $500$ from each chain. Convergence diagnostics for the MCMC samples for the selected model are provided in Section~S2.1.

\subsection{Specification of spatial covariates}\label{sec:model_selection_app}

For model selection, we first compared alternative variable selections and functional forms for the spatial covariates associated with the spatially varying coefficients in~\eqref{eq:svc}. This is achieved by employing a $10$-fold CV and analyzing the validation and goodness-of-fit $R^1(\tau)$ metric in~\eqref{eq:R1}. Each fold is trained on a dataset containing observations from $36$ stations, and $R^1(\tau)$ is evaluated on the remaining set of $4$ hold-out stations and averaged across stations. This model selection is carried out separately for the $0.50$ and $0.95$ quantiles.

We considered four alternative specifications (denoted $S_{1}$ to $S_{4}$), varying both the functional form and the inclusion of the distance-to-coast covariate. In all specifications, each spatial coefficient includes a global effect that centers the process, and the spatial intercept consistently incorporates the elevation effect. Specifications $S_1$ and $S_3$ restrict the distance-to-coast effect exclusively to the spatial intercept $\tilde{\beta}_0(\bs)$ using raw distance $\dist(\bs)$ and its logarithmic transformation $\log(\dist(\bs) + 1)$, respectively. In contrast, specifications $S_2$ and $S_4$ also incorporate $\log(\dist(\bs) + 1)$ into the remaining spatial coefficients, with $S_2$ keeping raw distance for the intercept and $S_4$ applying the logarithmic transformation to all coefficients. Specification $S_4$ corresponds to the actual full baseline specification in \eqref{eq:svc}. The formulations for these alternative specifications are detailed in Table~S1.

Table \ref{table:CV_spatial_struc_R1} shows the mean values of $R^1(\tau; \bs_i)$ for each climatic zone as well as the global mean of all stations for both study quantiles. Specification $S_4$ holds the best results of the metric $R^1(\tau)$ for both quantiles. Negative values for all specifications for the $0.50$ quantile are present in the CC, whereas negative values for the $0.95$ quantile are found for the MC. These results imply that the studied models perform worse than the baseline out of sample in some stations of these zones. However, it is worth noting that the $R^1(\tau)$ metric is a demanding metric, as its benchmark baseline relies on station-specific empirical quantiles. Furthermore, modeling coastal regions supposes a challenge since the lower thermal variability along the coast may contribute to lower $R^1(\tau)$ values, while complex sea-breeze dynamics and local marine microclimates may not be fully captured by the model predictors.

Individual results for each station may be found in Table~S2, where it can be appreciated that the best model may vary depending on the location, although, overall, $S_4$ holds the best results. Furthermore, Table~S3 highlights the clear advantage of utilizing a model specification that incorporates GPs ($S_4$) over a structurally analogous specification that omits them.

\begin{table}[tbp]
\centering
\scriptsize
\setlength{\tabcolsep}{4pt} 
\begin{tabular}{l *{8}{r}}
\toprule
& \multicolumn{4}{c}{$\tau = 0.50$} & \multicolumn{4}{c}{$\tau = 0.95$} \\
\cmidrule(r){1-1} \cmidrule(r){2-5} \cmidrule(l){6-9}
Zone / Specification & $S_1$ & $S_2$ & $S_3$ & $S_4$ & $S_1$ & $S_2$ & $S_3$ & $S_4$ \\
\cmidrule(r){1-1} \cmidrule(r){2-5} \cmidrule(l){6-9}
CC & $ -0.380 $ & $ -0.334 $ & $ -0.108 $ & $\mathbf{-0.023} $ & $ -0.166 $ & $ -0.134 $ & $ -0.021 $ & $\mathbf{0.042}$ \\ 
MC & $ 0.072 $ & $ 0.083 $ & $ 0.175 $ & $\mathbf{0.194} $ & $ -0.141 $ & $ -0.138 $ & $ -0.097 $ & $\mathbf{-0.085}$ \\ 
PC & $ 0.350 $ & $ 0.357 $ & $ 0.409 $ & $\mathbf{0.418} $ & $ 0.281 $ & $ 0.302 $ & $ 0.357 $ & $\mathbf{0.368}$ \\ 
\cmidrule(r){1-1} \cmidrule(r){2-5} \cmidrule(l){6-9}
Global mean & $ 0.171 $ & $ 0.185 $ & $ 0.273 $ & $ \mathbf{0.296} $ & $0.109 $ & $ 0.127 $ & $ 0.187 $ & $\mathbf{0.206} $ \\ \bottomrule
\end{tabular}%
\caption{Summary by climatic zone of the 10-fold CV $R^1(\tau)$ (for $\tau = 0.50$ and $\tau = 0.95$) across different spatial covariate specifications. The best results for each quantile are marked in bold.}
\label{table:CV_spatial_struc_R1}
\end{table}

\subsection{Assessment of long-term trends and atmospheric circulation contributions}
 
Choosing specification $S_4$ from the previous section as the best spatial covariate specification, we evaluate the contributions of the LT and AC of the model for both studied quantiles in the whole in-sample dataset.  We consider three model specifications. Model $M_1$ includes the spatial baseline, seasonal terms, and month-specific LT; model $M_2$ includes the spatial baseline, seasonal terms, and AC predictors; and model $M_3$ is the full model in \eqref{eq:model}, including both LT and AC predictors.

Table \ref{table:R1_M1M2M3} shows the mean $R^1(\tau;\bs_i)$ values across climatic zones and quantiles for the three models. Here, $R^1_{\text{LT}}(\tau)$, $R^1_{\text{AC}}(\tau)$, and $R^1(\tau)$ denote the marginal $R^1$ values for models $M_1$, $M_2$, and $M_3$, respectively, while $R^1_{\text{AC} \mid \text{LT}}(\tau)$ and $R^1_{\text{LT} \mid \text{AC}}(\tau)$ represent the partial $R^1$ metrics, which measure the predictive improvement of AC and LT conditional on the presence of the counterpart set of covariates, respectively. The results of model $M_2$ highlight the capability of the geopotential predictors when it comes to explaining the variability of the quantiles of daily maximum temperatures. The addition of the LT ($M_3$) improves the explained variability of the models; however, when only taking them into account ($M_1$), their explanatory power is weaker. Lower $R^1$ values along the coast may reflect the influence of local processes, such as sea-breeze dynamics and marine microclimates, that are not fully captured by the model predictors.
The partial $R^1$ values reveal a clear asymmetry in the predictive contribution of the covariates across both quantiles. Adding AC to a model already containing LT yields a substantial predictive improvement. In contrast, adding LT to a model that already accounts for AC provides a minor incremental gain. This is consistent with AC capturing the daily signal and part of the long-term variability, while LT accounts for the remaining long-term variability. These results are in line with station-wise $R^1(\tau;\bs_i)$ results (see Table~S4). 

\begin{table}
\centering
\scriptsize
\setlength{\tabcolsep}{4pt}
\resizebox{\textwidth}{!}{
\begin{tabular}{l rrr rr rrr rr}
\toprule
& \multicolumn{5}{c}{$\tau = 0.50$} & \multicolumn{5}{c}{$\tau = 0.95$} \\
\cmidrule(r){1-1} \cmidrule(r){2-6} \cmidrule(l){7-11}
Zone / $R^1$ & $R^1_{\text{LT}}(\tau)$ & $R^1_{\text{AC}}(\tau)$ & $R^1(\tau)$ & $R^1_{\text{AC} \mid \text{LT}}(\tau)$ & $R^1_{\text{LT} \mid \text{AC}}(\tau)$ & $R^1_{\text{LT}}(\tau)$ & $R^1_{\text{AC}}(\tau)$ & $R^1(\tau)$ & $R^1_{\text{AC} \mid \text{LT}}(\tau)$ & $R^1_{\text{LT} \mid \text{AC}}(\tau)$ \\ \cmidrule(r){1-1}
\cmidrule(r){2-6} \cmidrule(l){7-11}
CC & $ 0.125 $ & $ 0.227 $ & $ 0.237 $ & $ 0.124 $ & $ 0.012 $ & $ 0.038 $ & $ 0.178 $ & $ 0.183 $ & $ 0.150 $ & $ 0.007 $ \\ 
 MC & $ 0.205 $ & $ 0.263 $ & $ 0.305 $ & $ 0.124 $ & $ 0.056 $ & $ 0.120 $ & $ 0.194 $ & $ 0.229 $ & $ 0.123 $ & $ 0.043 $ \\ 
PC & $ 0.131 $ & $ 0.468 $ & $ 0.482 $ & $ 0.405 $ & $ 0.026 $ & $ 0.102 $ & $ 0.433 $ & $ 0.450 $ & $ 0.388 $ & $ 0.030 $ \\ \cmidrule(r){1-1} \cmidrule(r){2-6} \cmidrule(l){7-11}
 Global mean & $ 0.149 $ & $ 0.381 $ & $ 0.401 $ & $ 0.292 $ & $ 0.032 $ & $ 0.097 $ & $ 0.335 $ & $ 0.354 $ & $ 0.286 $ & $ 0.030 $ \\ \bottomrule
\end{tabular}
}
\caption{In-sample marginal and partial $R^1(\tau;\mathbf{s}_i)$ summary by climatic zone (for $\tau = 0.50$ and $\tau = 0.95$) across different model specifications.}
\label{table:R1_M1M2M3}
\end{table}

\begin{figure}[tb]
    \centering
    \includegraphics[width=0.85\linewidth]{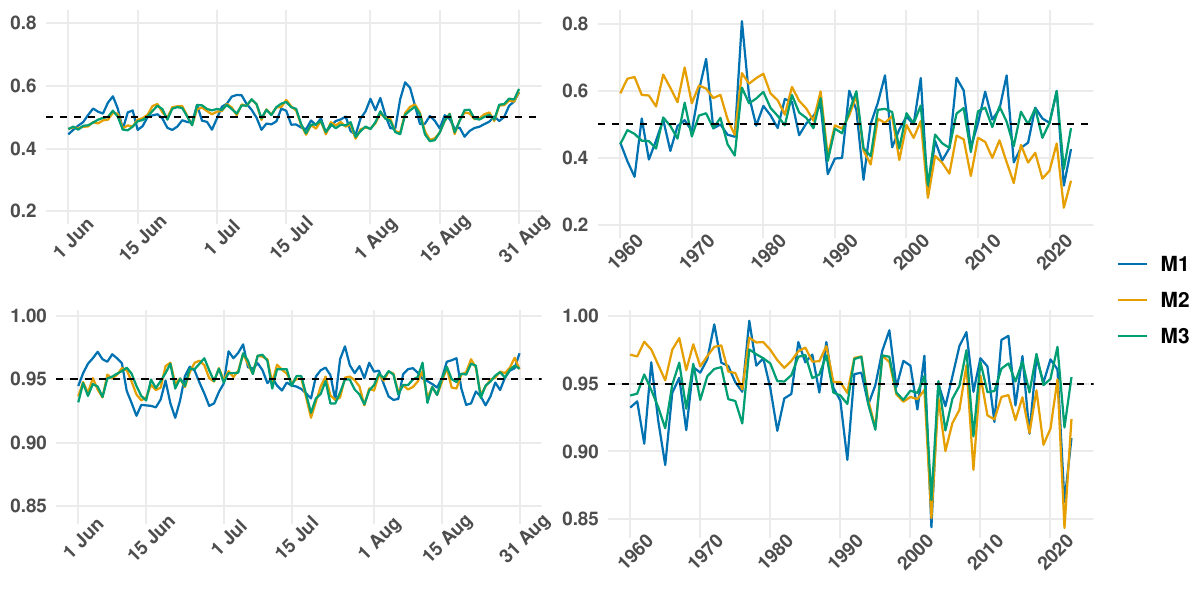}
    \caption{Empirical exceedance probabilities $p_\ell(\tau)$ (left column) and $p_t(\tau)$ (right column) in the 1960--2023 JJA period for $\tau = 0.50$ (top row) and $\tau = 0.95$ (bottom row).}
    \label{fig:p_t_l}
\end{figure}

Figure \ref{fig:p_t_l} shows the empirical exceedance probabilities $p_\ell(\tau)$ and $p_t(\tau)$ for $\tau = 0.50$ (top) and $\tau = 0.95$ (bottom) for the whole study period for both quantiles. The $p_t(\tau)$ results highlight the huge impact of adding LT terms in the model, making the models $M_1$ and $M_3$ capable of capturing the existing trend in both daily maximum temperature quantiles. On the other hand, results for $p_\ell(\tau)$ show that the seasonal terms included in all models satisfactorily capture the seasonal cycle. Additionally, the inclusion of AC terms in model $M_3$ is able to reduce the variability of the shown probabilities, making the results better centered around the studied quantiles. 
These results also hold when looking at station-wise probabilities $p_t(\tau;\bs_i)$ and $p_\ell(\tau; \bs_i)$ for both quantiles (see Figures~S34--S37). Finally, Table~S5 shows that every model performs satisfactorily when evaluating the station-wise probabilities $p(\tau;\bs_i)$, obtaining values very close to the studied quantiles.   

\subsection{Results for the selected model}
For the selected model from the previous section, $M_3$ with specification $S_4$ (same as the model presented in Section~\ref{sec:model}), we present a summary of the posterior distribution of the regression coefficients and parameters, and results obtained after the spatial interpolation. To present the results, we employ a fine grid $G_D$ of points within the study region. The grid contains $\vert G_D \vert = 4937$ grid cells, which correspond to a $10$ km $\times$ $10$ km resolution. Altogether, the entire posterior predictive dataset contains $64$ years $\times$ $92$ days $\times$ $4937$ grid centroids $\times 1000$ replicates $\approx 29.1$ billion points. 

\subsubsection{Model parameters}

\begin{figure}[tb]
    \centering
    \includegraphics[width=\linewidth]{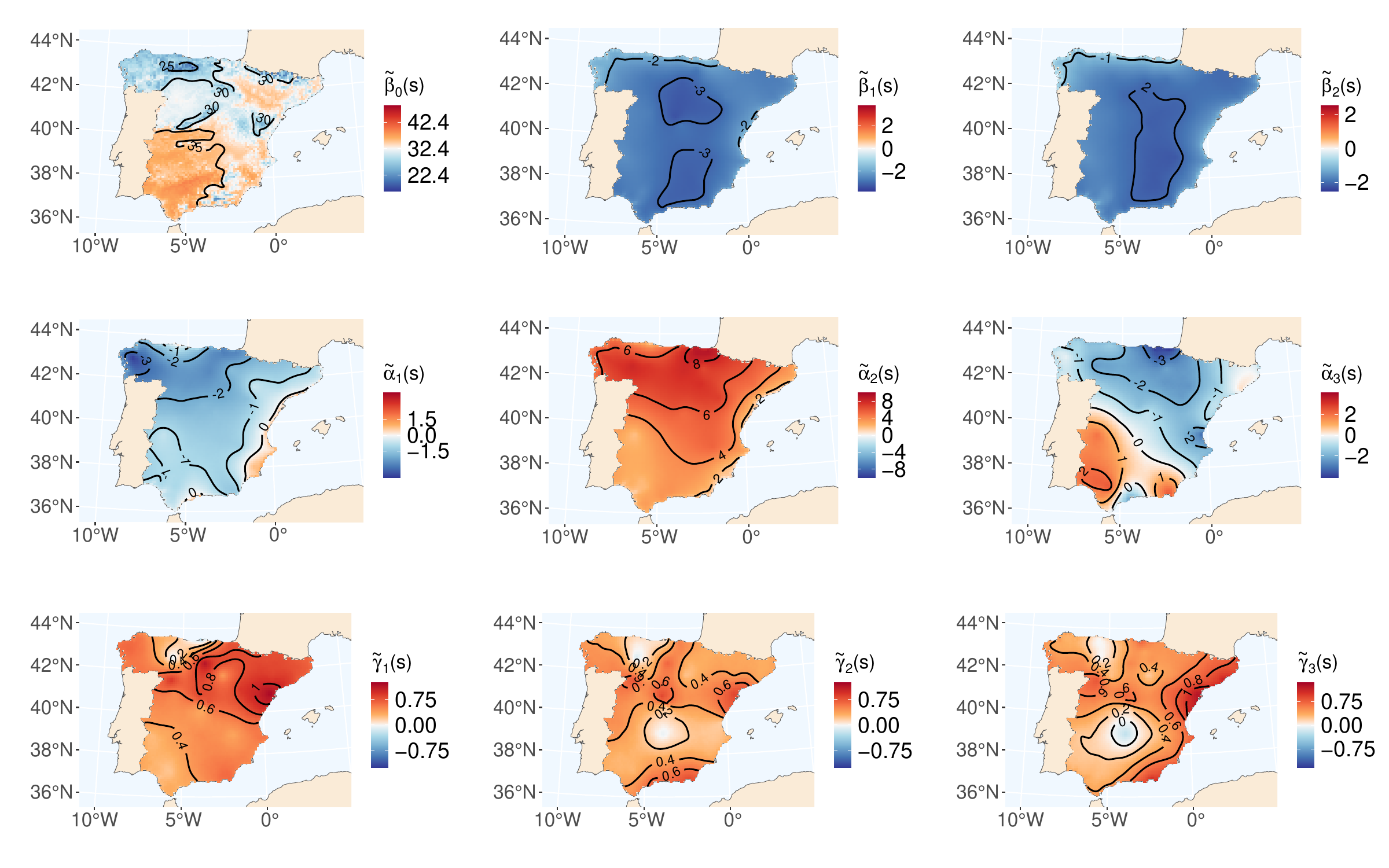}
    \caption{Posterior mean surfaces for the spatial processes centered on an intercept and corresponding spatial variables in the $0.95$ quantile model across the region of study. Top: Intercept and seasonal terms. Mid: AC. Bottom: LT.}
    \label{fig:kriging_q0.95}
\end{figure}

Figure~\ref{fig:kriging_q0.95} shows the posterior mean surfaces for the spatial coefficients defined in \eqref{eq:svc} for $\tau = 0.95$. These surfaces allow us to assess the local effects of the covariates across the study region when taking into account the geography.
The spatial intercept $\tilde\beta_0(\bs)$ is the only process that incorporates the elevation, making its effect perceptible by finding elevated regions with clearly lower effect magnitudes with respect to their surroundings. The rest of the covariates only incorporate the logarithm of the distance to the coast plus one. The spatial processes corresponding to the seasonal effects, $\tilde\beta_1(\bs)$ and $\tilde\beta_2(\bs)$, exhibit negative effects, especially in the center of the region of study. The AC spatial processes present different effects. For the geopotential anomaly at pressure level $300$ hPa, its process $\tilde\alpha_1(\bs)$ presents mainly negative effects, with weaker effects along the coasts, especially the MC. On the other hand, the process $\tilde{\alpha}_2(\bs)$ corresponding to the geopotential anomaly at pressure level $500$ hPa exhibits important positive effects, with the greatest values along the CC. The pressure level $700$ hPa coefficient $\tilde{\alpha}_3(\bs)$ shows mixed results, with positive values in the southwest and negative values towards the northeast. Finally, looking at the surfaces for the LT for June ($\tilde\gamma_1(\bs)$), July ($\tilde\gamma_2(\bs)$), and August ($\tilde\gamma_3(\bs)$), we can appreciate clearly positive effects, which imply an additional long-term warming in the $0.95$ quantile of daily maximum temperature. These effects vary depending on the month, having different effects along both coasts and the center of the peninsula. In conclusion, these results highlight the different effects of the geography for each predictor. 
Results for the quantile $\tau = 0.50$ can be found in Figure~S38, and results for the posterior mean surfaces of the GPs centered at zero for the $0.50$ and $0.95$ quantiles can be found in Figures~S39 and S40, respectively. 

Next, we look at the posterior mean and posterior $95\%$ CIs of the remaining model parameters. 
The residual scale parameter $\sigma$ for the $0.50$ and $0.95$ quantile models has a posterior mean and $95\%$ CI of $0.962$ $(0.958, 0.966)$ and $0.251$ $(0.250, 0.252)$, respectively. Posterior means and $95\%$ CI for the GPs hyperparameters may be found in Table~S6, highlighting the difference in the standard deviation between processes. The results of the posterior distributions of the effective ranges suggest a configuration in which the effective spatial ranges are physically reasonable with the dimensions of the study region.

\subsection{Inference results}

This section provides inference results for the selected model to assess the different contributions to climate change in the $0.50$ and $0.95$ daily maximum temperature quantiles during the JJA period in $1960$--$2023$ across the study region. Specifically, we quantify the overall warming while separating the individual contributions arising from shifting AC patterns and LT.

\subsubsection{Monthly difference between 32-year periods} 

\begin{figure}[tb]
    \centering
    \includegraphics[width=\linewidth]{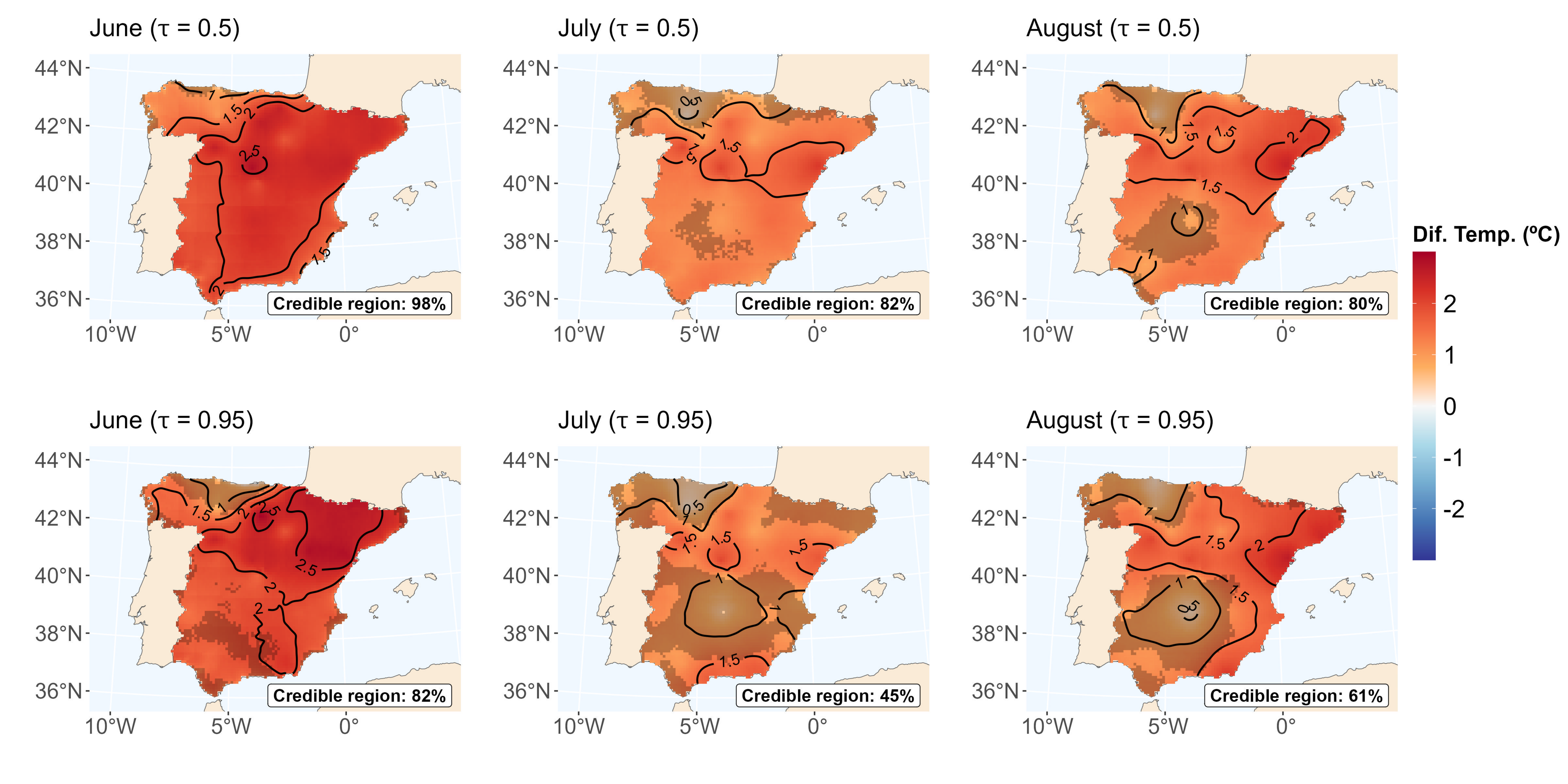}
    \caption{Mean monthly differences between the last ($1992$--$2023$) and first ($1960$--$1991$) 32-year periods observed for the $0.50$ (top) and $0.95$ (bottom) quantiles. The grid cells whose $95\%$ CI contain zero are shaded.}
    \label{fig:dif_decades}
\end{figure}

To observe the warming of the predicted quantiles, we compute the monthly mean difference between the last (1992--2023) and the first (1960--1991) 32-year periods observed, separately for each quantile. To give uncertainty information about the differences, we use the MCMC samples to obtain $95\%$ CIs. Under the hypothesis of climate change, these differences should be positive, and the $95\%$ CIs should not contain zero in an important part of the region of study.  Figure~\ref{fig:dif_decades} shows the results obtained, with shaded grid cells as the locations where the $95\%$ CI contains the zero value. For the $0.50$ quantile, approximately $98\%$, $82\%$, and $80\%$ of the region exhibits credibly positive differences, and mean and $95\%$ CI of $2.00$ ($1.04, 2.49$) $^\circ$C, $1.26$ ($0.57, 1.82$) $^\circ$C, and $1.39$ ($0.74, 2.10)$ $^\circ$C for the months of June, July, and August, respectively. In particular, we find that the northern part of the PC and the MC are highly affected by climate change, with a greater effect than in the CC. 
In the case of the $0.95$ quantile, June, July, and August present credibly positive differences in approximately $82\%$, $45\%$, and $61\%$ of the region, with means and $95\%$ CI equal to $1.98$ ($0.92, 2.72$) $^\circ$C, $1.13$ ($0.47, 1.66$) $^\circ$C, and $1.37$ ($0.55, 2.23$) $^\circ$C, respectively. The most relevant warming is found in June in the MC, Iberian System, and Central System. These results show evidence of climate change that is different in space and time.

\begin{figure}[tb]
    \centering
    \includegraphics[width=\linewidth]{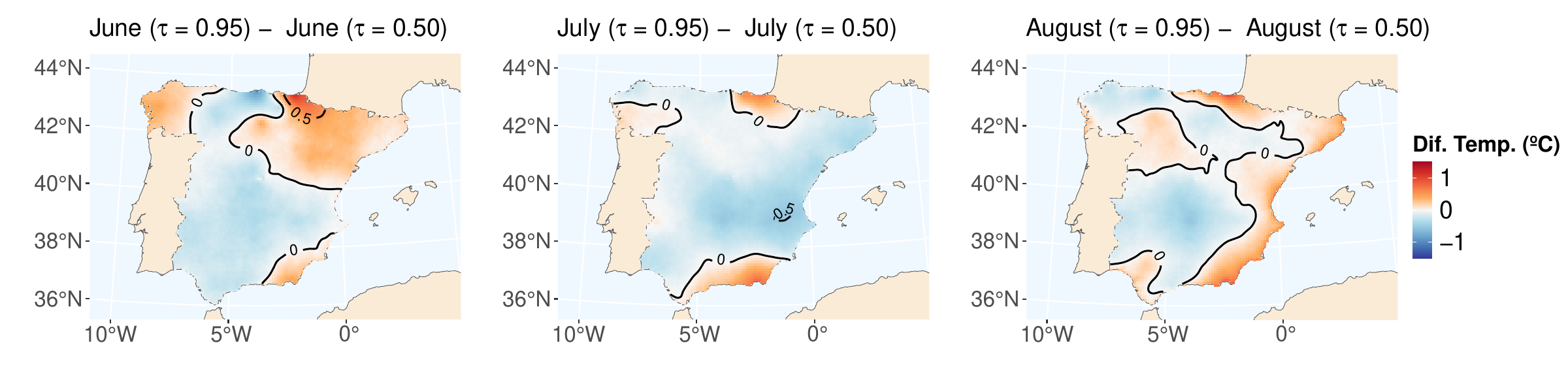}
    \caption{Mean monthly differences between the $0.95$ and $0.50$ quantile warming between the last ($1992$--$2023$) and first ($1960$--$1991$) 32-year periods observed.}
    \label{fig:dif_decades_quantiles}
\end{figure}

Additionally, the differences between the 0.95 and 0.50 quantiles of the monthly differences computed above are examined to assess how warming varies across quantiles and to identify areas where the differences between these quantiles are greatest. Figure~\ref{fig:dif_decades_quantiles} highlights the spatio-temporal heterogeneity between quantiles, despite the absence of substantial global differences, with mean values of $-0.03^\circ$C, $-0.13^\circ$C, and $-0.02^\circ$C  for June, July, and August, respectively. In June, the $0.95$ quantile has suffered greater warming in the northeastern part of the Peninsula, with a warming close to $0.5^\circ$C with respect to the $0.50$ quantile. In July, the warming of the $0.50$ quantile has been greater in the majority of the region. Lastly, in August, the $0.95$ quantile is clearly higher along the MC and the Pyrenees, whereas the $0.50$ quantile has undergone greater warming in the central part. 

Comparing the last (2014--2023) and first (1960--1969) observed decades highlights a stronger climate change effect. Regional warming exceeds $3.5^\circ\text{C}$ in specific areas, while local differences between quantiles sometimes exceeds $0.5^\circ\text{C}$ in absolute terms (see Figures~S41 and S42).

\begin{figure}[tb]
    \centering
    \includegraphics[width=\linewidth]{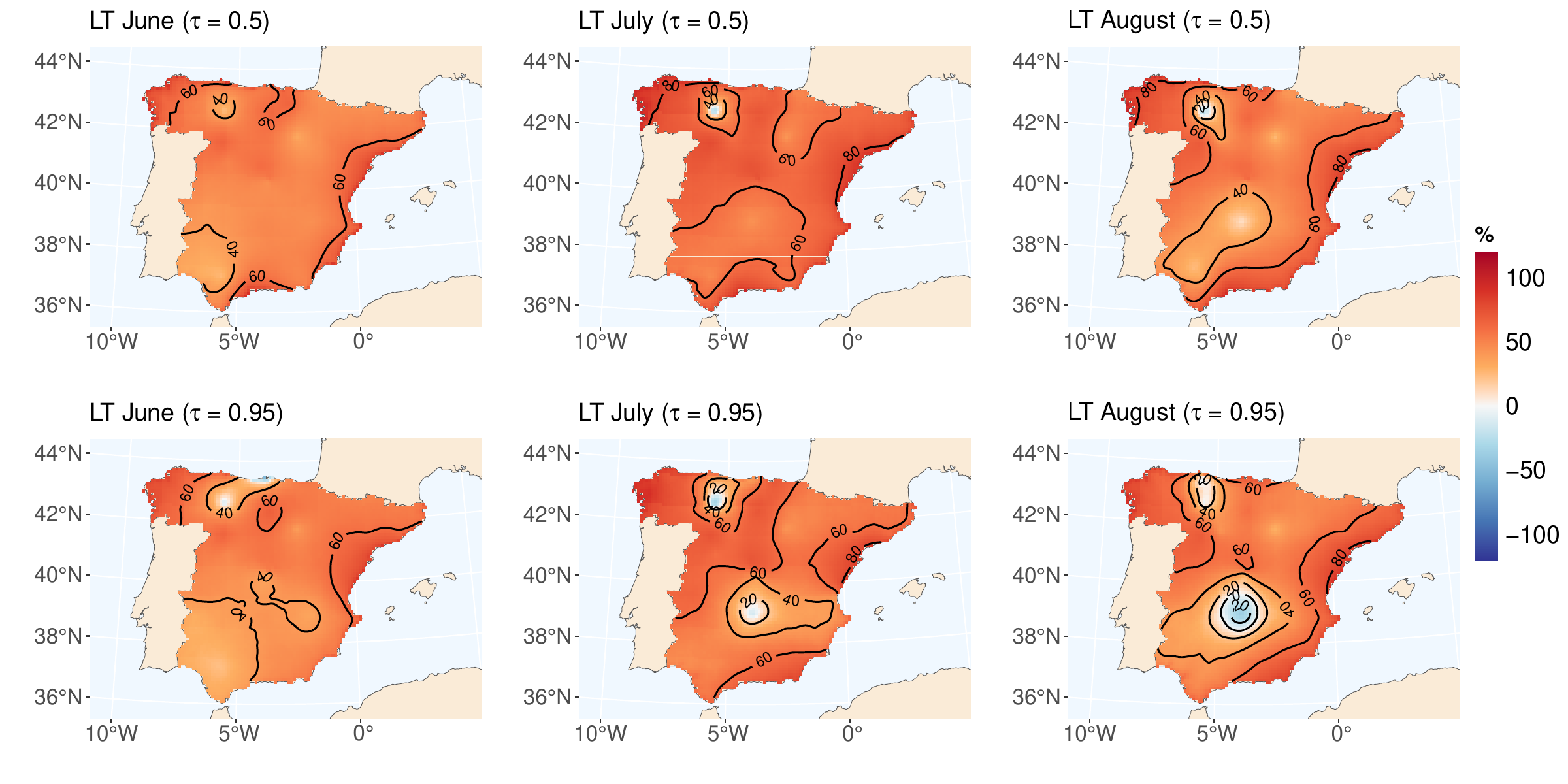}
    \caption{Monthly long-term trend percentage contribution relative to the mean quantile warming between the last ($1992$--$2023$) and first ($1960$--$1991$) 32-year periods observed for $\tau = 0.50$ and $\tau = 0.95$.}
    \label{fig:perc_LTT}
\end{figure}

Lastly, thanks to the additive structure of our model, we can directly decompose the mean differences between decades into the relative percentage contributions of the LT and AC (see Section~S2.5.2). Figure~\ref{fig:perc_LTT} shows the LT percentage contribution relative to the mean quantile warming between the last and first 32-year periods observed, with the AC contribution accounting for the remaining percentage. 
Incorporating spatially varying coefficients for all predictors allows us to capture the complex spatio-temporal heterogeneity of both contributions. Across all months and quantiles, the MC consistently exhibits LT contributions above $50\%$ (frequently reaching $60$--$80\%$), highlighting the strong dominance of background thermodynamic warming along the eastern coast. Localized negative LT values indicate instances where AC was the primary driver of warming, offsetting a cooling trend component. Finally, a comparison between quantiles reveals that the $0.50$ quantile displays a smoother spatial distribution with peak LT dominance during July, whereas the $0.95$ quantile exhibits greater spatial variability. Globally, the percentage contribution of LT for the $0.50$ quantile presents a mean and $95\%$ CI of $51.29$~($33.3, 74.77$)~\%, $63.48$~($47.65, 85.69$)~\%, and $52.63$~($26.10, 81.99$)~\%, for June, July, and August, respectively. For the $0.95$ quantile, we find values of $48.42$~($26.24, 73.97$)~\%, $54.94$ ($18.35, 81.23$)~\%, and $49.20$~($-4.60, 79.72$)~\% for June, July, and August, respectively. Figure~S43 presents the results when comparing the last and first decades observed.

\subsubsection{Block average across years}

Furthermore, we evaluate the spatial block average \citep[Chapter 7]{banerjee2025} of the $\tau$ quantile over a subregion $K \subseteq D$ for day $\ell$ within year $t$ defined as
\begin{equation*}
     Q_{Y_{t\ell}(K) \mid \mathbf{G}_{t\ell}(\mathbf{b})}(\tau) = \frac{1}{\lvert K \rvert} \int_K Q_{Y_{t\ell}(\bs) \mid \mathbf{G}_{t\ell}(\mathbf{b})} (\tau) \, d\bs.
\end{equation*}
In practice, this integral is approximated by using a grid $G_K = G_D \cap K$ and averaging over the grid points $\bs_k \in G_K$, i.e.,
\begin{equation*}
     Q_{Y_{t\ell}(K) \mid \mathbf{G}_{t\ell}(\mathbf{b})}(\tau) \approx \frac{1}{\lvert G_K \rvert}\sum_{\bs_k \in G_K} Q_{Y_{t\ell}(\bs_k) \mid \mathbf{G}_{t\ell}(\mathbf{b})} (\tau).
\end{equation*} 

\begin{figure}
    \centering
    \includegraphics[width=0.75\linewidth]{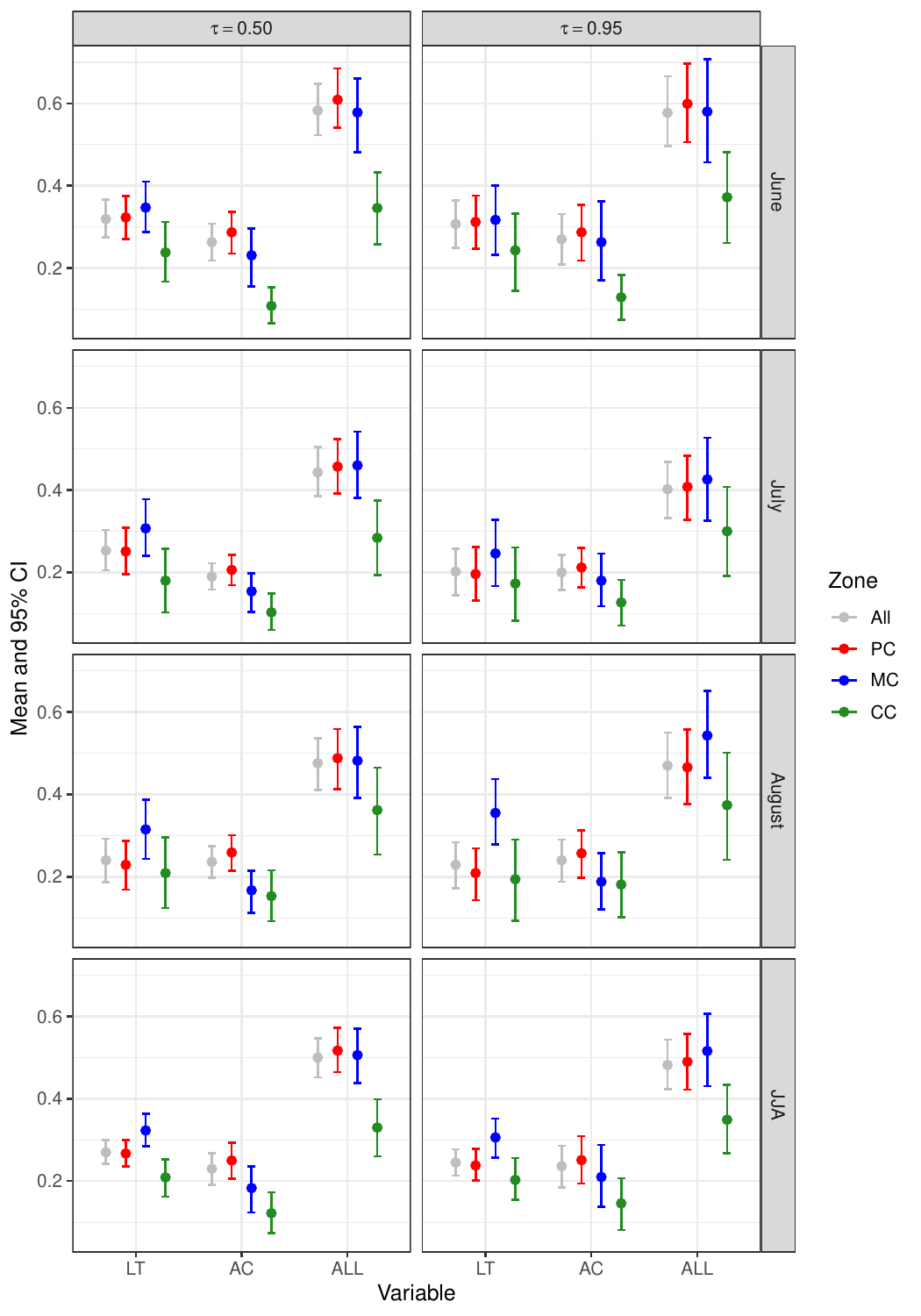}
    \caption{Decadal mean slopes ($^\circ$C/decade) and $95\%$ CIs derived from linear models fitted to yearly spatial block-averages of the isolated components (LT and AC), and the full model (All) across MCMC iterations, broken down by quantile, period, and subregion.}
    \label{fig:bck_avg_coef}
\end{figure}

Analyzing this block average across years provides insight into the regional temporal evolution of the studied quantile. Moreover, in the proposed spatio-temporal model, the isolated contributions of LT and AC can be assessed by setting the non-target component to zero. 
Figure~\ref{fig:bck_avg_coef} summarizes the decadal mean slopes ($^\circ$C/decade) and $95\%$ CIs obtained by fitting an ordinary least squares linear model to the yearly block-averages across all MCMC iterations, broken down by quantile, month, and climatic zone (see Table~S7 for numerical results).

Across both quantiles, clear regional disparities emerge. In particular, the CC consistently exhibits lower slopes, with numerous $95\%$ CIs that exclude the point estimates of the other zones. Regarding the driver components, the contribution of the LT is generally larger than the AC contribution in most cases.  However, their relative importance is significantly different in some cases, such as the results for the MC in August for both quantiles, where the point estimates for the LT contributions are not contained in the $95\%$ CIs of the other zones.
 Lastly, consistent behavior across quantiles is found, except for the MC in August, where the $0.95$ quantile displays a substantially steeper trend ($0.54^\circ$C/decade) than the $0.50$ quantile ($0.48^\circ$C/decade), in line with the difference between quantile warmings found in Figure~\ref{fig:dif_decades_quantiles}.

\section{Summary and future work}\label{sec:conclusions}

In this work, we have proposed a spatially varying quantile downscaling model to assess the contribution of large-scale AC and LT to local summer temperature. We combine daily maximum temperatures with geopotential fields from ERA5, which provide a description of large-scale synoptic atmospheric configuration. We considered three tropospheric levels, 300, 500, and 700 hPa on a $1^\circ \times 1^\circ$ spatial grid, which capture different aspects of the atmospheric structure. We assumed that the response of temperature to a given atmospheric configuration varies across the domain, reflecting regional differences in how synoptic circulation translates into surface temperature. Additionally, the proposed model includes LT, which allows the evolution of temperature to differ across months, representing the temporal change in temperature that is not explained by variations in AC. The proposed methodology has been applied to analyze the $0.50$ and $0.95$ quantiles of daily maximum temperatures for $64$ years during the summer months in peninsular Spain. 

Results reveal heterogeneous evidence of climate change in daily maximum temperatures across time, space, and quantiles. The mean difference between the last (1992--2023) and first (1960--1991) 32-year periods observed presents positive values for both quantiles, with $95\%$, $82\%$, and $80\%$ of the region exhibiting credibly positive differences for June, July, and August, respectively, for the $0.50$ quantile, and $82\%$, $45\%$, and $61\%$ for the $0.95$ quantile. The northeastern part of the peninsula is the most affected, with quantile warmings above $2^\circ$C across the summer period between these 32-year periods. Between-quantile differences between the previous 32-year periods show that, although the spatial mean difference is near zero, we find local differences near $\pm 0.5^\circ\text{C}$. In particular, in August, the $0.95$ quantile has had a greater warming along the MC than the $0.50$ quantile ($0.54$ ($0.44, 0.65$) vs $0.48$ ($0.39, 0.56$) $^\circ\text{C}/\text{decade}$). The relative contribution of LT and AC in these mean differences also presents heterogeneity across space, time, and quantiles. Notably, LT consistently dominates along the MC, accounting for over $60$\%--$80$\% of the difference across both quantiles and all summer months. Lastly, evaluating spatial block-averages of both quantiles across zones reveals differences in the temporal evolution of the $0.50$ and $0.95$ quantiles. 

While this study relies on ERA5 daily geopotential height fields at three pressure levels over a spatial grid, the proposed framework is flexible and generalizable. The methodology can readily be transferred to other geographical domains using different global or regional reanalysis products, regardless of whether they feature finer or coarser spatial resolutions. Furthermore, the model architecture is not restricted to geopotentials; it can accommodate a broader set of atmospheric or surface predictors depending on the specific study objective and physical mechanisms governing temperatures in the target region.

It is worth noting that, although the proposed approach accounts for quantile-specific variability, it can also be readily adapted to mean regression by replacing the AL errors with Gaussian errors. This alternative may be particularly appealing when the model already explains most of the variability in the data and the primary interest lies in the conditional mean rather than in the distributional features captured by different quantiles.

While the present study focuses on a descriptive characterization, extending this framework to predictive contexts, such as the projection of future temperature quantiles under CMIP6 climate scenarios, represents a key direction for future research.  In such predictive settings, navigating the high dimensionality of atmospheric covariates across multiple pressure levels becomes critical. Incorporating spatial Bayesian variable selection into our GP framework offers a promising avenue to identify the dominant physical drivers while fully preserving spatial dependence and uncertainty quantification. By delivering fine-scale projections of extreme heat events, this methodology will provide crucial insights to inform local public health interventions and regional environmental mitigation strategies.

\section*{Supplementary information}

The online version contains supplementary material available. 

\section*{Acknowledgements}

This work was partially supported by MCIN/AEI/10.13039/501100011033 and Unión Europea NextGenerationEU under Grant PID2023-150234NB-I00, and BIOSTATNET network RED2024-153680-T; and Gobierno de Aragón under Grants PROY\_T21\_24, UNIZAR 15075, and Research Group `Modelos Estocásticos'. J. M.-G. was supported by Gobierno de Aragón under Doctoral Scholarship ORDEN EMC/590/2025.

\section*{Declarations}
\begin{itemize}
\item \textbf{Competing Interests:} The authors have no competing interests to disclose.)
\end{itemize}

\bibliography{bibliography}

\end{document}